\documentclass[aaspp4]{aastex}
\shorttitle{Merging Galaxies in the SWIRE Fields}
\shortauthors{Huang \& Hwang}

\begin{document}
\title{Morphologically-Identified Merging Galaxies in the SWIRE Fields}
\author{Mei-Ling Huang and Chorng-Yuan Hwang}
\affil{Graduate Institute of Astronomy, National Central University, Jungli, Taoyuan 320,
Taiwan}
\email{mlhuang@astro.ncu.edu.tw; hwangcy@astro.ncu.edu.tw}

\begin{abstract}     %Abstract
We investigate the evolutional and environmental effects on star
formation efficiency for more than 400 merging galaxies. The
$\sim$400 merging systems, with photometric redshifts smaller than
0.7, are obtained from a catalog of $\sim$15000
morphologically identified merging galaxies derived from
observations of the Canada-France-Hawaii Telescope. We also obtained
the IR data of the merging galaxies from the {\it Spitzer} Wide-area
InfraRed Extragalactic Survey (SWIRE). The redshift differences
$\Delta z$ between the member galaxies of these merging pairs
show a large distribution with $0 < \Delta z < 0.4$.
We divide our merging pairs into two sub-samples with
$\Delta z < 0.05$ and $> 0.05$ for
further analyses. We find a statistically significant
anti-correlation between the specific star formation rate
(SSFR) and the separation of the merging galaxies for both sub-samples.
Our analyses also show that although most of the merging systems do have
enhanced star formation activity, only very rare ones display extremely
high SFRs. Additionally, the SSFR of the merging galaxies also
decreases when the magnitude difference between two member galaxies
becomes large. However, we find that for the merging pairs with
large luminosity contrast, the fainter components show
higher SSFR than the brighter ones. Finally, there is a higher
fraction of gas-poor mergers in galaxy clusters, and the SSFR of
gas-rich mergers is reduced in cluster environments.
\end{abstract}

%Keyword

\section{Introduction} % Sec 1
Galaxy interaction and merging play an essential role in galaxy
formation and evolution. Numerical simulations showed that merging
processes might transform blue late-type galaxies into red
early-type galaxies \citep{tis02}. It was also predicted that tidal
interactions between galaxy pairs could transform spiral and
irregular galaxies into S0 and elliptical galaxies
\citep{too72,mih96,bar96}. Some observations showed that
ellipticals formed through the merger of spirals \citep[e.g., the
Antennae;][]{sch82,hib96}. Non-axisymmetric gravitational potential
arising during the galaxy interactions might produce tidal torques
that can induce strong gas inflows, which might trigger an intense
burst of star formation. Galaxy mergers are therefore well known to
be responsible for triggering enhanced star formation activity
\citep[e.g.,][]{ken98,bar00,lam03}.

An interesting issue of merging galaxies is how the triggered star
formation activity is related to the internal physical parameters of
the galaxy mergers, such as pair separations and mass ratios. It has
been reported that star formation rate (SFR) is anti-correlated with
the pair projected separation and radial velocity
\citep[e.g.,][]{bar00,lam03}. Since SFR scales with galaxy mass,
it is better to use the specific star formation rate (SSFR),
defined as the SFR divided by the stellar mass, to denote star
formation activity of galaxy \citep[e.g.,][]{nik04}. \citet{nik04}
discovered that the SSFR also shows an anti-correlation with the
projected separation. In
addition, the induced star formation activity is also related to the
mass ratio of the galaxy pair; the pairs with commensurate
luminosities trigger stronger star formation activity than those
with large luminosity contrast \citep{lam03,woo06}. In the minor
mergers, \citet{lam03} found that brighter members show a higher
probability to have tidally enhanced star formation than the fainter
ones, but \citet{woo07} reported that the fainter galaxies show
higher SSFRs than the brighter ones.

Besides the internal parameters, the triggered star formation
activity might also be affected by environments. \citet{alo04} found
that the pairs in clusters required a smaller separation than those
in the fields to show obvious enhanced star formation activity by
studying 2-degree field (2dF) pairs. They also pointed out that the
pairs in dense environments showed lower enhanced star formation
activity than those in the fields. Nevertheless, it is difficult to
determine whether this SFR inhibition is true or mainly results from
the effect of the morphology-density relation \citep{dre80} since
ellipticals, which are dominant in clusters, have much smaller star
formation activity than spirals.

Early studies based on UV/optical observations were susceptible to
dust extinction
\citep[e.g.,][]{bar00,lam03,nik04,alo04,woo06,woo07}. During the
merging process the gas inflows induced by tidal torques of galaxy
interaction are converted into new stars with large amounts of dust,
which is heated by the young stars and radiates energy at
mid-infrared (MIR) to far-infrared (FIR) wavelengths; this indicates
that infrared (IR) is an ideal tool to study mergers.

Furthermore, previous studies on the relevant topics were all based on
merging galaxies selected using the inter-galaxy separations. Such
methods would tend to select interacting galaxies at relatively
early stages of the merging processes. To statistically study the
influence of the merging process and environments, we need a large
un-biased merging galaxy sample based on consistent searching
criteria. In this study, we select a sample of
morphologically-identified merging galaxies \citep{hwa09}, which are
not prone to specific merging stages, and we combine the infrared
and optical data to investigate the properties of these merging
galaxies.

This paper is organized as follows. Section 2 introduces our data
and data processing. In Section 3, we discuss the photometric
redshifts of the galaxy pairs and estimate the star formation rates.
In Section 4, we study the distribution of SSFR in the merging
galaxies. The relationships of the SSFR with the projected
separation and the mass ratio of the galaxy pairs are analyzed in
Section 5. In Section 6 we discuss the environmental effects on the
SSFR of the mergers. Finally, Section 7 summarizes our main
findings. Where needed, we assume a standard $\Lambda$CDM cosmology
with $\Omega_{M}$ = 0.3, $\Omega_{\Lambda}$ = 0.7, and $H_{0}$ = 70
km s$^{-1}$ Mpc$^{-1}$ throughout this study.

\section{Data and Data Reduction}   %  Sec 2
\subsection{Sample of Merging Galaxies}     %2.1
The merging galaxies in this study are obtained from a catalog of
interacting and merging galaxies derived from the Red Sequence
Cluster Survey 2 (RCS2) of Canada-France-Hawaii Telescope
observations \citep{hwa09}. This catalog is the largest catalog of
galaxy-merging systems assembled to date and includes more than
15,000 interacting and merging galaxies via morphological
classification. Possible candidates were first selected by an
automatic morphological-pattern-recognition package and subsequently
inspected by human eyes to assure their credibility in the final
catalog. These sources constitute a reliable and uniform sample of
merging galaxies for further study. In addition, the RCS2 data used
to develop this catalog covers extensive and separate areas up to
422 deg$^{2}$, providing a great opportunity to conduct
multi-waveband researches combined with other large surveys, such as
the {\it Spitzer} Wide-area Infrared Extragalactic Legacy Survey
\citep[SWIRE;][2004]{lon03} and the Sloan Digital Sky Survey
\citep[SDSS;][]{yor00}.

The SWIRE is the largest extragalactic {\it Spitzer} Legacy Science
programs, imaging $\sim$49 deg$^{2}$ at the wavelengths of Infrared
Array Camera (IRAC) 3.6, 4.5, 5.6, 8~$\mu$m and Multiband Imaging
Photometer (MIPS) 24, 70, 160~$\mu$m. The SWIRE project has mapped
seven high-latitude fields; the RCS2 regions are overlapped with
3 of these 7 fields---ELAIS N1, ELAIS N2, and Lockman Hole. These 3
SWIRE fields are listed in Table 1. We note that these fields have
also been specifically selected to be far away from the ecliptic to
avoid possible contaminations from wandering asteroids as well.
Within these 3 fields there are 540 merging systems identified in
the catalog of \citet{hwa09}.

\subsection{Optical Data from SDSS}     % 2.2 SDSS

In order to have accurate photometric information for these
morphologically identified merging galaxies, we also obtain the
optical data from the SDSS survey. The SDSS is a combined imaging
and spectroscopic survey of 10$^{4}$ deg$^{2}$ in the north Galactic
cap and a smaller region in the south. The SDSS imaging gathers data
in five broad bands, $ugriz$, ranging from 3000 to 10,000 \AA. We
use the imaging data in the seventh SDSS data release (DR7) in this
study.

We first retrieve a list of galaxies within an angular distance of
10$\arcsec$ from the 540 merging systems. We visually examine each
image of the 540 merging systems to identify their SDSS
counterparts. It turns out that a small part of the merging systems
are identified as foreground stars by SDSS. We find 60 merging
systems composed of at least one component identified as stars by
SDSS; we reject these sources from our sample. We then classify the
remaining 480 merging systems, totaling to 723 galaxies, into 3
groups according to the number of their members: merger (MG; 1),
close pair (CP; 2), and close multiple (CM; more than three). The
final sample consists of 253 MGs, 213 CPs, and 14 CMs.
In the following analyses, we only focus on the MGs and CPs
to study the relations between the SSFR and some physical parameters,
such as pair separation, pair magnitude difference, and the
environments. The CMs are ignored in these analyses
because they are complicated and too difficult to quantify
the effects of the physical parameters.
Some examples of these merging systems are shown in Figure 1.

The SDSS database contains various measured magnitudes for each
detected source; we select model magnitudes for our targets. We also
acquire the photometric redshifts for our galaxies because of their
lack of spectroscopic observations. Notwithstanding the accuracy of the
photometric redshifts is not as good as that of spectroscopic
redshifts, the former still provides sufficient information for
extragalactic and cosmological researches, especially in the
statistical study of the evolutionary properties for faint objects.

\subsection{IR Data from SWIRE}     % 2.3 SWIRE

We search the IR counterparts of the merging galaxies in the SWIRE
fields to study their properties of star formation activity. We
mainly utilize the SWIRE data at 3.6 and 24~$\mu$m. The 3.6~$\mu$m
luminosity can be used as a tracer of the stellar component since
the NIR emission is dominated by older stellar populations and less
susceptible to recently formed massive stars
\citep[e.g.,][]{wu05,dav06,li07,han07,RR08}. On the other hand, the
MIR emission can be used to estimate the total IR luminosity for a
variety of galaxies \citep{cha01,elb02,pap02,tak05}. We use the 24~$\mu$m
flux to derive the total IR luminosities and thus the SFRs of
the merging galaxies.

\subsubsection{IR Counterparts from SWIRE Catalogs}

We obtain the IR data for bright sources from the SWIRE catalogs. We
first check the SWIRE flag---extended source flag---for each source
in SWIRE catalogs to exclude ``definitely point-like'' objects. The
processed SWIRE catalogs are subsequently cross-correlated with our
723 merging galaxies; the IR counterparts are determined by
identifying the closest SWIRE sources within an angular distance of
5$\arcsec$ from each merging galaxy. We find 482 and 152 galaxies
with IR counterparts at 3.6 and 24~$\mu$m respectively. The SWIRE
catalogs provide several kinds of flux measurements; for our
objects, we select the Kron flux \citep{kro80}, which is especially
preferable for extended sources.

\subsubsection{IR Counterparts from SWIRE Images}

The SWIRE catalogs only contain sources with sufficient fluxes. For
example, the threshold for the detection at 3.6~$\mu$m required
SNR~$\geq 10$, which is equivalent to 10~$\mu$Jy for almost all of
the survey area. To retrieve faint sources which were not included
in the SWIRE catalogs, we extract the fluxes within apertures of
1.9$\arcsec$ ($\sim$1.6 FWHM) and 5.25$\arcsec$ ($\sim$1.9 FWHM)
centered at the position of each merging galaxy in the
IRAC-3.6~$\mu$m and MIPS-24~$\mu$m images separately. These images
are downloaded from the NASA/IPAC Infrared Science Archive and have
been post-processed and co-added. They are the same ones used to
make the SWIRE catalogs. We obtain the IR fluxes of the merging
galaxies if the corresponding extracted objects have an SNR greater
than 3~sigma. Combining the faint sources and the objects from the
SWIRE catalogs, we find 600 and 621 out of 723 merging galaxies with
IR counterparts at 3.6 and at 24~$\mu$m respectively. There are
fewer galaxies with IR counterparts at 3.6 than at 24~$\mu$m because
some of the galaxies are out of the IRAC imaging fields.

Since we can not detect very faint sources at high redshifts, we
might have some biases in selecting the merging pairs. For examples,
if we divide our sample into major merging pairs with $\Delta m_{z}
< 1.5$ and minor merging pairs with $\Delta m_{z} \geq 1.5$, we
might tend to detect the minor mergers at relatively low redshifts
becasue we can have more detectable low-luminosity sources at lower
redshifts. Nevertheless, the redshift distributions of the major and
minor mergers are not distinct; the mean redshift for the major
mergers is $0.17 \pm 0.11$ and that for the minor mergers is $0.15
\pm 0.10$. This indicates that this flux-limited bias is negligible.

\section{Sample Characteristics}        % Sec 3
\subsection{Photometric Redshifts of Close Pairs}  %3.1
The redshift distribution of our sample is shown in Figure 2. The
farthest merging galaxies could reach out to redshifts as far as
$z\sim0.7$; the observed 24~$\mu$m flux for this largest redshift
corresponds to the rest-frame 14.1~$\mu$m flux, still within the MIR
range. Figure 3 shows the distribution of the redshift difference,
$\Delta z$, between two member galaxies for each CP. At first
glance, this wide distribution might imply that our sample is
contaminated by spurious galaxy pairs; i.e., some pair
galaxies might not be really close in physical space. However, we
note that our sample is selected through morphological
identification; most CPs exhibit clear features of merging galaxies,
such as ``tail'' and ``bridge''. Some examples of the CPs with
$\Delta z > 0.05$ are shown in Figure 4. As can be seen, these CPs
do show the features of merging galaxies in spite of their large
redshift difference. We thus conclude that the redshift differences
for most of these CPs are not caused by projection.

The discrepancy of photometric redshifts in some of the CPs is not
surprising. Enhanced star formation in the interacting galaxies may
affect their color caused by strong line emission and dust
absorption in the star forming region. This would influence their
photometric redshift estimates derived by fitting observational data
with synthetic templates. We have considered different criteria in
obtaining a better redshift estimate between two interacting galaxies,
such as selecting the photometric redshift from the brighter galaxy
or selecting the smaller redshift. We find that our derived results
are all statistically consistent in spite of the selecting criteria
used; one of the main reasons is that the redshift effects are
substantially canceled out when we consider the SSFRs derived from
the ratios of luminosities (fluxes) of different wavebands. For
simplicity, we assume the smaller photometric redshift between two
member galaxies as the true redshift for those CPs with $\Delta z >
0.05$ in presenting our results.

\subsection{Estimating Total IR Luminosity and Stellar Mass }  % 3.2

Emission of MIR is an ideal tracer of the total IR luminosity.
Previous studies have developed libraries of luminosity- or
color-dependent galaxy templates to calculate the total IR
luminosity from 24~$\mu$m flux densities
\citep{dal01,cha01,dal02,lag03}. We derive the total IR luminosities
from the 24~$\mu$m flux densities adopting the templates from
\citet{cha01}; these templates consist of a combination of 105
spectral energy distributions (SEDs) from normal to starburst
galaxies.

We employ the SDSS $z$-band ($\lambda_{eff}$ = 9097 \AA) luminosity as
the stellar mass indicator. The 3.6~$\mu$m flux density was usually
used as an estimate for the stellar mass
\citep[e.g.,][]{wu05,dav06,li07,han07,RR08}.
As found by \citet{cha96} and \citet{mad98}, the mass-to-NIR ratio,
independent of either galaxy color or Hubble type, is less susceptible
to the star formation history and the same for all types of galaxies.
Nonetheless, there are 123 sources in our sample without information
at 3.6~$\mu$m due to
detection limit or to the lack of IRAC observations. On the other
hand, it has also been reported that $z$-band luminosity can be used
as an un-biased estimate of the stellar mass \citep{nik04}. To check the
credibility of using the SDSS $z$-band as the stellar mass indicator
in our sample, we plot the 3.6~$\mu$m versus the SDSS $z$-band
luminosity for the 600 sources with both 3.6~$\mu$m and $z$-band
detections in Figure 5. The 3.6~$\mu$m and the SDSS $z$-band
luminosities are calibrated to the rest frame using the templates
from \citet{cha01}. As can be seen, the 3.6~$\mu$m luminosity is
proportional to the SDSS $z$-band luminosity. The $z$-band luminosity
provides a reliable estimate of stellar mass. We have hence
obtained an estimate of the stellar mass based on the SDSS $z$-band
luminosity for every galaxy in our sample.

Throughout this study, we use $L_{IR}/L_{z}$ to represent the SSFR.
The SFR scales with the galaxy mass and might fail to
reflect the star formation efficiency; in contrast,
the SSFR, which has taken into account the effect of stellar mass,
is a more suitable estimate of star formation activity in galaxies.
The $L_{IR}/M_{*}$ ratio has been employed to represent the SSFR
in previous studies \citep[e.g.,][]{lin07}.
They used rest-frame (B-V) colors and absolute $B$-band
magnitude to derive stellar mass. We note that the $B$-band magnitude is
susceptible to dust extinction and requires careful calibration as done
by \citet{lin07}. In this paper, we use rest-frame SDSS $z$-band
luminosity to represent the stellar mass because the SDSS $z$-band
luminosity is proportional to the rest-frame IRAC 3.6~$\mu m$
luminosity (Figure 5), which is less susceptible to star formation history
and dust extinction and is a better mass tracer.
For each CP, we use $L_{IR}/L_{z}$ to represent the total SSFR by
summing the total IR luminosity of both galaxies over
the total $z$-band luminosity of both members.
We note that the distance influence on the SSFR is small because
the division of $L_{IR}$ by $L_{z}$ will have eliminated most of
the distance effects. Thus, the uncertainties in the photometric
redshift estimates are not as vital as might have been expected.

\section{SSFR Distribution of Galaxy Pairs } % Sec 4
%\subsection{Bimodal Distribution of SSFR}           % 4.1

We first explore characteristic features of the SSFR distribution of
the CPs. Upper panel of Figure 6 shows the SSFR distribution of 213
CPs. This distribution shows a significant dip at
$L_{IR}/L_{z}\sim1-1.5$, which looks like a boundary of two
different distributions. To understand the origin of this boundary,
we adopt the SED templates from \citet{pol07} to generate the
$L_{IR}/L_{z}$ values for various types of galaxies. Table 2
summarizes the values of $L_{IR}/L_{z}$ for a variety of galaxies
ranging from ellipticals to ultra-luminous infrared galaxies
(ULIRGs). As can be seen in Table 2, while the galaxies with
$L_{IR}/L_{z}$ below unity are normal ones, the galaxies with
$L_{IR}/L_{z}$ larger than unity all exhibit enhanced star formation
activity. This indicates that the bimodal distribution of Figure 6
actually symbolizes the galaxies with and without enhanced star
formation activity. Some galaxies of the CPs are influenced by tidal
interactions and manifest enhanced star formation activity whereas
others still remain unaffected.

We try to find suitable distribution functions to describe the SSFR
distribution. First, the CPs on the left side looks like a normal
distribution, which seems to be reasonable for normal galaxies. The
probability density function (p.d.f.) of a normal distribution is

\begin{equation}
f(x\mid \mu, \sigma) = \frac{1}{\sigma\sqrt{2\pi}} e^{\frac{-(x-\mu)^2}{2\sigma^{2}}},
  \label{eq:1}
\end{equation}

where $\mu$ and $\sigma$ are the mean and standard deviation of the
variable $x$. On the other hand, we might expect the distribution of
galaxies with enhanced star formation activities to be lognormal;
that is, only very few sources have extremely star formation
activities. The right hand side of the SSFR distribution shows a
concentration of the $L_{IR}/L_{z}$ values at 1.5--2 and a long tail
toward the upper end, also corresponding with the
characteristics of a lognormal distribution. The p.d.f. of a
lognormal distribution is

\begin{equation}
f(x\mid \mu, \sigma) = \frac{1}{x\sigma\sqrt{2\pi}} e^{\frac{-(lnx-\mu)^2}{2\sigma^{2}}},
 \label{eq:2}
\end{equation}

where $\mu$ and $\sigma$ are the mean and standard deviation of the
variable's natural logarithm. The lognormal distribution is closely
related to normal distribution. If $x$ is distributed lognormally
with parameters $\mu$ and $\sigma$, $log(x)$ is distributed normally
with the mean $\mu$ and standard deviation $\sigma$. We adopt a normal
function to fit the left part and a lognormal distribution function
to fit the right part of the SSFR distribution. Table 3 lists the
derived parameters, and the results are plotted in the middle panel
of Figure 6.

The lognormal fits are somehow unsatisfactory. This is due to the
fact that at the low end there are always galaxies with ``normal''
star formation efficiency, and on the high end, there are more
galaxies than model predicted. In order to correct these
shortcomings, we construct a modified lognormal p.d.f. :

\begin{equation}
f(x\mid \mu, \sigma) =(x-1)^{3} \cdot \frac{1}{x\sigma\sqrt{2\pi}} e^{\frac{-(lnx-\mu)^2}{2\sigma^{2}}}.
  \label{eq:3}
\end{equation}

We note that such modification is purely from mathematical
reasoning. The derived parameters are also listed in Table 3, and
the fitting results are plotted in Figure 6 (lower panel). The data
are better fitted with a mixture of the normal and the modified
lognormal distribution functions. We note that the modified
lognormal distribution is still very close to the standard lognormal
distribution. We conclude that the SSFRs of the CPs that remain
unaffected by tidal interactions are normally distributed, but the
SSFRs of those with enhanced star formation will approximately
follow a lognormal distribution.

\section{Dependence of SSFR on Physical Parameters} %5
\subsection{Projected Separation}           %5.1

We explore effects of the projected separation $\Delta r_{p}$ of the
merging galaxies on the SSFR. We ignore the CPs  without 24~$\mu$m
detection; the total remained CPs in this analysis is 181. Figure 7
shows the dependence of $L_{IR}/L_{z}$ on the projected separation;
the circle and plus signs represent the CPs with $\Delta z \leq
0.05$ and $\Delta z > 0.05$ respectively, and the triangle
represents the MGs. In Figure 7, the $L_{IR}/L_{z}$ shows a
generally declining envelop as a function of $\Delta r_{p}$. To test
the statistical significance of this relation, we compute their
correlation coefficients and probabilities. The linear correlation
coefficient $r$ for the overall sample is $-0.18$, with the
probability of the correlation $P = 0.98$, indicating $L_{IR}/L_{z}$
and $\Delta r_{p}$ are significantly anti-correlated. The sample
with $\Delta z \leq 0.05$ shows a correlation of $r = -0.27$ with $P
= 0.94$, and the sample with $\Delta z > 0.05$ shows a correlation
of $r = -0.16$ with $P = 0.94$. To reduce the effect of
outliers in the correlation test, we also apply the Spearman rank
method, which is more robust than the linear correlation. A Spearman rank
correlation test also confirms the anti-correlation between the SSFR
and the projected separation; we obtain $r_{s} = -0.17$ and $P_{s} = 0.97$
for the overall sample, $r_{s} = -0.33$ and $P_{s} = 0.98$ for the
sample with $\Delta z \leq 0.05$, and $r_{s} = -0.17$ and $P_{s} =
0.95$ for the sample with $\Delta z > 0.05$. All these tests show
that the anti-correlation between the SSFR and the observed
projected separation $\Delta r_{p}$ of merging galaxies is
statistically significant.

Our results qualitatively agree with previous studies. It was
reported that SSFR is anti-correlated with pair projected separation
for the merging galaxies of local universe \citep{nik04}.
\citet{lin07} also found a declining envelop of $L_{IR}/M_{*}$ as a
function of projected separation for pairs at higher redshift (z
$\sim$ 0.1--1.1); however, no statistical analysis was provided and
is difficult to evaluate and to compare with their results
quantitatively.

In our sample, the CPs with similar projected separations spread a
wide range of the $L_{IR}/L_{z}$ values. We note that CPs with
similar projected separations might in fact have very diverse real
separations; this could cause large scattering in the $L_{IR}/L_{z}$
values. CPs with similar physical separations but
undergoing different merging stages would also contribute to the
scatter of SSFR because some CPs may just start to approach, while
others might have experienced merging processes more than once and
had drained most of the gas in the early processes. Likewise,
the same fact might explain why the $L_{IR}/L_{z}$ values of MGs
also scatter over a wide range.
It was also noted that the SSFR is related to the galaxy types or intrinsic
properties of the merging galaxies \citep{lam03,nik04} and the
luminosity contrast (or mass contrast) between member galaxies.
Given all these uncertainties, our results still show a significant
anti-correlation between the SSFR and the observed projected
separation $\Delta r_{p}$ of merging pairs; this demonstrates that the
internal separations between merging galaxies must have strong
influence on the SSFRs.

\subsection{Mass Ratio}         % 5.2

We investigate the dependence of the SSFR on the mass ratio of our CPs.
We use the $z$-band magnitude difference between the pair galaxies,
$\Delta m_{z}$, to represent their mass ratio. We divide our sample
into major merging pairs with $\Delta m_{z} < 1.5$ and minor merging
pairs with $\Delta m_{z} \geq 1.5$. The number of the major merging
pairs is 134, including 38 CPs with $\Delta z \leq$ 0.05 and 96 CPs
with $\Delta z > 0.05$; the 47 minor merging pairs contain 11 with
$\Delta z \leq 0.05$ and 36 with $\Delta z > 0.05$. The distribution
of the $z$-band magnitude differences for the CPs is shown in Figure
8.

We first examine the relation between $L_{IR}/L_{z}$ and $\Delta
r_{p}$ to test the influence of the luminosity contrast for the
major and minor merging pairs. Figure 9 shows
$L_{IR}/L_{z}$ versus $\Delta r_{p}$ of these two types of merging
pairs. We find a significant anti-correlation between the
$L_{IR}/L_{z}$ and $\Delta r_{p}$ for the major merging pairs with
$r_{s} = -0.20$ and $P_{s} = 0.98$ using the Spearman rank test. We
further divide the major merging pairs into two subgroups with
$\Delta z \leq 0.05$ and $\Delta z > 0.05$ separately. The
statistical significance of the anti-correlations of these two
subgroup are $r_{s} = -0.43$ and $P_{s} = 0.99$ for the major
merging pairs with $\Delta z \leq 0.05$ and $r_{s} = -0.18$ and
$P_{s} = 0.92$ for the ones with  $\Delta z > 0.05$ respectively. In
other words, the correlation between $L_{IR}/L_{z}$ and $\Delta
r_{p}$ becomes weaker for the major merging pairs with $\Delta z
> 0.05$;  this might be caused by the errors of the true distance
determination for some sources in this sample.

In contrast, the minor merging pairs display no correlations
between $L_{IR}/L_{z}$ and $\Delta r_{p}$. The Spearman test show
$r_{s} = -0.08$ and $P_{s} = 0.41$ for the minor merging pairs.
There are no anti-correlations for both subgroups with $\Delta z
\leq 0.05$ and $\Delta z > 0.05$ either. This is consistent with the
results of \citet{woo06} and implies that the intrinsic properties
of the minor merging galaxies might be dominant factors in
determining the SSFR of the minor merging galaxies.

We next study the relation between the SSFR and the magnitude
difference in these merging galaxies pairs. Figure 10 shows the
relation between $L_{IR}/L_{z}$ and $\Delta m_{z}$ for our sources.
The results of the Spearman rank test exhibit a clear
anti-correlation between the $L_{IR}/L_{z}$ and $\Delta m_{z}$ with
$r_{s} = -0.21$ and $P_{s} = 0.99$ for the overall sample; the
correlations are slightly weaker with  $r_{s} = -0.30$ and $P_{s} =
0.97$ for the pairs of $\Delta z \leq$ 0.05 and $r_{s} = 0.16$ and
$P_{s} = 0.93$ for the pairs of $\Delta z > 0.05$. This
anti-correlation between the $L_{IR}/L_{z}$ and $\Delta m_{z}$ is
also consistent with the result of \citet{woo06}.

We also compare the SSFRs of the faint galaxies and the bright
galaxies in the minor merging pairs. To estimate the stellar mass
 of galaxies, we apply the relation
between the $z$-band and 3.6~$\mu m$ luminosities of our sample as
shown in Figure 5,
\begin{equation}%========= Eq 4===============
log\left(\frac{L_{3.6\mu m}}{L_{\odot}}\right)=(0.17\pm0.13)+(0.92\pm0.01)log\left(\frac{ L_{z}}{L_{\odot}}\right),
\label{eq:4}
\end{equation}
and the relation found by \citet{li07},
\begin{equation}%========= Eq 5 ===============
log\left(\frac{M_{*}}{M_{\odot}}\right)=(1.34\pm0.09)+(1.00\pm0.01)log\left(\frac{\nu L_{\nu}[3.6 \mu m]}{L_{\odot}}\right).
\label{eq:5}
\end{equation}  %========= Eq 5===============
We note that equation (4) also holds in different redshift ranges,
indicating that its redshift dependence is small.

Dwarf galaxies are excluded in calculating the SSFRs since the
dwarves might have depleted their gas and their SSFRs are not
affected by tidal interaction anymore. We exclude dwarf galaxies with
$\log(M/M_{\odot}) < 9$, which corresponds to galaxies about one
order magnitude fainter than the characteristic
luminosity of field galaxies \citep{bla05}.

We calculate the individual SSFRs of the faint galaxies and the
bright galaxies in the minor merging pairs. As shown in Figure 11,
all the faint galaxies have higher SSFRs than the bright galaxies.
Their average SSFRs are listed in Table 4; the average SSFRs of
the major merging galaxies are also listed as a comparison.
Although the major merging pairs have higher
SSFRs than the minor pairs, the average SSFR of the faint galaxies
in the minor merging pairs is much higher than that of the major
merging pairs. This is due to the fact that the SSFRs of the bright
galaxies in minor merging pairs are much lower than those of the
major merging pairs; the very low SSFRs and the high mass of the
bright galaxies compensate the high SSFRs of the faint galaxies in
the minor merging pairs. As a result, the major merging pairs have
higher SSFRs than the minor ones on average. Our results indicate
that the SSFR of faint galaxies are more susceptible to the effects
of tidal interaction.

This result is in agreement with that of \citet{woo07} but are
inconsistent with that of \citet{lam03}. \citet{lam03} found that in
the minor merging pairs, the effects on star formation activities
are more important in the brighter galaxies. We note that their
faint galaxy sample including sources down to $\log(M/M_{\odot}) \sim
8$, which might contain many dwarf galaxies. These dwarf galaxies
might have depleted most of their gas and could not trigger any new
star formation even with very strong tidal interaction. This shows
that the intrinsic properties of galaxies have to be taken into
account when we consider the effect of tidal interaction in a
merging pair.

\subsection{Estimation of Superposition}  %%=== 5.3 =====

The redshift differences $\Delta z$ between the member galaxies of
our merging pairs show a large distribution with $0 < \Delta z < 0.4$.
It is difficult to directly quantify the contamination
because our sources have been specifically selected via morphological
pattern recognition. This selection criterion has largely removed
superposed false pairs. Nonetheless, we might evaluate the
effects of overlapping pairs on our results from some correlation tests.
We note that the internal separations between merging galaxies have a
strong correlation with the SSFRs \citep[e.g.,][]{nik04}, so the
correlation results could reflect the reliability of our sample.
We divide the pairs with $\Delta z > 0.05$ into three subgroups
according to their redshift differences and calculate
the Spearman rank correlation test on the SSFR and the separation
for each subgroup, as listed in Table 5.

We find that there are no correlations between $L_{IR}/L_{z}$ and
$\Delta r_{p}$ for CPs with $0.05 < \Delta z < 0.2$
but a significant one for CPs with $\Delta z > 0.2$
($r_{s} = -0.24$ and $P_{s} = 0.98$). In Figure 3, we find a small
bump around $0.1 < \Delta z < 0.2$ and the distribution becomes flat
for $\Delta z > 0.2$. From the correlation tests, we speculate that
the bump around $\Delta z \sim 0.1$ to 0.2 might be caused by superposition,
whereas most of the sources with  $\Delta z > 0.2$ are actually true
pairs with wrong redshift determination. This unusual result is caused by
the fact that there are only a few high-redshift sources and thus
the probability of superposition involving one member at high
redshifts is low. It turns out that there are very few overlapping
pairs with true large $\Delta z$.
From the bump of Figure 3, we estimate that there are about fifty
contaminating sources by comparing the distribution of large $\Delta z$ sources.

\section{ENVIRONMENTAL EFFECTS ON SSFR} %  Sec 6
\subsection{Environmental Distribution of Wet/Dry/Mixed Merging Pairs} % 6.1

Environments also have strong influence on the merging galaxies. To
study the environmental effects on galaxy evolution, we compare the
distribution of merging galaxy types in the fields with that in the
clusters. The cluster regions for comparison are selected from the
maxBCG cluster catalog \citep{koe07}, which covers similar regions
of our sample galaxies and is the largest sample of observed galaxy
clusters available by far. The criteria of assigning a CP in a
cluster are (1) the redshift difference between the CP and its host
cluster is smaller than 0.05, and (2) the projected separation
between the CP and the center of the cluster is smaller than 1000
kpc.

We classify the types of galaxies according to their SSFRs. A galaxy
is classified as a red (or dry) galaxy if $L_{IR}/L_{z} < 0.5$ and a
blue (or wet) one if $L_{IR}/L_{z} > 0.5$.  Since some galaxies have
no 24~$\mu$m detection with 3$\sigma$ upper limit,
 the galaxies with the $L_{IR}/L_{z}$ upper limit lower than 0.5
are also classified as red galaxies. There are about 16 sources,
which have the $L_{IR}/L_{z}$ upper limits larger than 0.5 and are
difficult to classify. We could either assign these uncertain sources
as blue galaxies or red galaxies. In the former case, we would have the
upper limit for the total number of the blue galaxies and in the
latter case the upper limit for the red ones. The results from the
former and latter classifications are represented as class 1 and
class 2 in Table 7 separately.

As can be seen in Table 7, the dry-to-wet ratio in the cluster is
higher than that in the field. The dry-to-wet ratio of class 1
serves as a lower limit of the ratio because those galaxies with the
$L_{IR}/L_{z}$ upper limit higher than 0.5 might actually be red
galaxies. On the other hand, the dry-to-wet ratio in class 2 is the
upper limit one as we classify all galaxies with the $L_{IR}/L_{z}$
upper limit into red galaxies. Hence, the dry-to-wet ratio in the
clusters is around 0.25--0.75 and the ratio in the field is around
0.03--0.09. In addition, only about 3\% of all the wet merging
galaxies are located in clusters, but about 24--29\% of dry merging
pairs and 18--21\% of mixed merging pairs are in clusters. Dry
mergers show a higher tendency to populate the clusters than other
types of mergers; this tendency might be related to the
morphology-density relation in which early-type galaxies
preferentially populate high density environments while late-type
galaxies are likely to inhabit low density environments \citep{dre80}.

\subsection{Inhibition of Star Formation Activity in the Cluster} % 6.2
The influence of cluster environment on the SSFRs of galaxies is
obvious for our merger sources (MGs). These interacting sources have
merged together and the individual galaxies can not be identified.
We find that the average SSFR of the MGs in the fields (156 sources)
is $2.47\pm0.20$, and the average SSFR of those in the clusters (9
sources) is $1.86\pm0.41$. We note that, however, we could
not exclude the effects of the morphology-density relation
\citep{dre80} on the different SSFRs of these MGs since we do not
know the galaxy types of the progenitors of these mergers.

In order not to confuse the environmental effects with intrinsic
pair properties, we further select only the wet major merging pairs
both in the clusters and in the fields to explore whether their
SSFRs are different. We use $L_{IR}/L_{z}=0.5$ as the dividing point
to separate the wet galaxies from the dry ones. The result is shown
in Table 8.

We find that the SSFRs of the galaxy pairs are reduced in the
clusters. The average SSFR of the wet major merging galaxies in the
clusters is 2.28 and that in the fields is 4.36. We note that the
average separation of the pairs in the clusters is 13.29 kpc,
similar to that of pairs in the field, 13.14 kpc; this suggests that
the separation effect is not important. Besides, since we consider
only wet major merging galaxies, the different SSFRs can not be
associated with the morphology-density relation as in \citet{alo04}.

Furthermore, we find an anti-correlation between the SSFR and $L_{z}$
among the MGs in the fields (Figure 12). The Spearman rank test shows a
highly significant result with $r_{s} = -0.23$ and $P_{s} = 0.99$,
indicating that in the fields the enhanced star formation activity
of the more massive MGs is weaker than that of the less massive
ones. Nonetheless, there is no correlation for the MGs in the clusters
with $P_{s} < 0.50$. This implies that the star formation activity
and/or merging history are very different for the MGs in the fields
and those in the clusters.

Our results demonstrate that the environments of clusters do inhibit the
star formation activity among galaxy pairs. This inhibition might be
due to (1) ram pressure stripping of cold interstellar medium of
galaxies falling into the cluster through the intracluster medium
\citep[e.g.,][]{gun72,nul82,fuj99,aba99}; and/or (2) strangulation,
which means that the diffuse gas in galaxy halos is stripped by the
gravitational potential of the cluster, cutting off the supply of
cold gas \citep[e.g.,][]{lar80,bek02}. On the other hand, for the
MGs in the fields, the mass of the MGs is anti-correlated with the
SSFR. This might be due to the fact that the more massive MGs have
undergone more merging processes than the less massive ones and have
exhausted more gas and dust in the early precesses.

\section{CONCLUSION} % Sec 7

We assemble a sample of more than 400 merging systems to study
their features of star formation efficiency. Our sample is drawn
from a catalog of $\sim$15000 morphologically identified merging
galaxies. We identify their IR counterparts in the SWIRE data
and their optical counterparts in the SDSS database. Statistical
analyses are performed to explore the SSFR distribution, the
relations between the SSFR and the pair separation and magnitude
difference, and the environmental effects on the SSFR. We summarize
our main results in the following:

1. The SSFR distribution of the merging galaxies follows a bimodal
distribution. The SSFRs of the galaxy pairs that remain unaffected
by tidal interactions have a normal distribution; on the other hand,
the SSFRs of the pairs with enhanced star formation activity
approximately follow a lognormal distribution. In other words,
although most merging systems do show enhanced star formation
activity, merging systems with extremely high star formation rates
are rare.

2. There is an anti-correlation between the SSFR and the projected
separation of the merging galaxies. The SSFR of the pairs also
declines when the magnitude difference between two member galaxies
increases. The pairs with comparable luminosities produce stronger
star formation activity in total than those with large luminosity
contrast; however, in a minor merging pair, the fainter component
show significantly higher SSFR than the brighter one.

3. We find a higher ratio of dry to wet mergers in galaxy clusters
than that in the field. Dry mergers show a higher tendency to
populate the clusters than other types of mergers. In addition, the
SSFRs of wet major mergers are also inhibited in the clusters.

\acknowledgements

This work was partially supported by the National Science Council
through grant NSC~99-2112-M-008-014-MY3 and NSC~99-2119-M-008-017.
This research has made use of the NASA/ IPAC Infrared Science
Archive, which is operated by the Jet Propulsion Laboratory,
California Institute of Technology, under contract with the National
Aeronautics and Space Administration. Funding for the SDSS and
SDSS-II has been provided by the Alfred P. Sloan Foundation, the
Participating Institutions, the National Science Foundation, the
U.S. Department of Energy, the National Aeronautics and Space
Administration, the Japanese Monbukagakusho, the Max Planck Society,
and the Higher Education Funding Council for England. The SDSS Web
Site is http://www.sdss.org/. The SDSS is managed by the
Astrophysical Research Consortium for the Participating
Institutions. The Participating Institutions are the American Museum
of Natural History, Astrophysical Institute Potsdam, University of
Basel, University of Cambridge, Case Western Reserve University,
University of Chicago, Drexel University, Fermilab, the Institute
for Advanced Study, the Japan Participation Group, Johns Hopkins
University, the Joint Institute for Nuclear Astrophysics, the Kavli
Institute for Particle Astrophysics and Cosmology, the Korean
Scientist Group, the Chinese Academy of Sciences (LAMOST), Los
Alamos National Laboratory, the Max-Planck-Institute for Astronomy
(MPIA), the Max-Planck-Institute for Astrophysics (MPA), New Mexico
State University, Ohio State University, University of Pittsburgh,
University of Portsmouth, Princeton University, the United States
Naval Observatory, and the University of Washington. Parts of this
research are based on observations obtained with MegaPrime/MegaCam,
a joint project of CFHT and CEA/DAPNIA, at the Canada-France-Hawaii
Telescope (CFHT) which is operated by the National Research Council
(NRC) of Canada, the Institute National des Sciences de l'Univers of
the Centre National de la Recherche Scientifique of France, and the
University of Hawaii. Access to the CFHT was made possible by the
Ministry of Education and the National Science Council of Taiwan as
part of the Cosmology and Particle Astrophysics (CosPA) initiative.

\clearpage
%%Table

\begin{deluxetable}{cccc}
\tablecaption{SWIRE Fields Overlapped
with RCS2}\label{tbl1}
\tablewidth{0pt}
\startdata \hline\hline
Field  &\multicolumn{2}{c}{Center Coordinate (J2000)}                 &   Area (deg$^{2}$)        \\
            &R.A.                       & Decl.     &               \\ \hline
ELAIS N1         &$16^{\mathrm{h}}11^{\mathrm{m}}00^{\mathrm{s}}$   &$+55^{\mathrm{d}}00^{\mathrm{m}}00^{\mathrm{s}}$&  9.00 \\
ELAIS N2         &$16^{\mathrm{h}}36^{\mathrm{m}}48^{\mathrm{s}}$   &$+41^{\mathrm{d}}01^{\mathrm{m}}45^{\mathrm{s}}$&  4.45    \\
Lockman Hole    &$10^{\mathrm{h}}45^{\mathrm{m}}00^{\mathrm{s}}$    &$+58^{\mathrm{d}}00^{\mathrm{m}}00^{\mathrm{s}}$&  14.32\\
\enddata
\end{deluxetable}

\begin{deluxetable}{ll}
\tablecaption{$L_{IR}/L_{z}$ of various types of galaxies based on
models in \citet{pol07}}
\tablewidth{0pt}
\tablehead{ \colhead{Galaxy type} & \colhead{$L_{IR}/L_{z}$}
}
\startdata
E           & 0.11-0.13\\
S0          & 0.17\\
Sa              & 0.25\\
Sb          & 0.42\\
Sc          & 0.56\\
Sd          & 1.03\\
Sdm             & 1.08\\
NGC6090 (SB)    & 17.25\\
M82 (SB)    & 20.62\\
Arp220 (ULIRG)  & 29.57\\
\enddata
\end{deluxetable}

\begin{deluxetable}{lcc}    % Table 3
\tablecaption{Parameters of the fitting results}\label{tbl3}
\tablewidth{0pt}
\tablehead{ \colhead{p.d.f.} & \colhead{$\mu$}  & \colhead{$\sigma$}}
\startdata
normal          & 0.65      &  0.34     \\
lognormal       & 0.70      &  0.43      \\ \hline
normal          & 0.65      &  0.34     \\
modified lognormal  & $-7.50$   &  1.14      \\
\enddata
\tablecomments{Here $\mu$ and $\sigma$ are the fitting parameters for the SSFR distribution of galaxy
pairs in the forms of the normal and lognormal distribution functions. For the normal function,
$\mu$ and $\sigma$ are the mean and standard deviation of the SSFR. For the lognormal function,
$\mu$ and $\sigma$ are the mean and standard deviation of the natural logarithm of the SSFR.
}
\end{deluxetable}

\begin{deluxetable}{lcc}  % Table 4
\tablecaption{Average SSFRs of bright and faint galaxies in minor mergers and all galaxies in major mergers}\label{tbl4}
\tablewidth{0pt}
\tablehead{  \colhead{CP}  & \colhead{member}   & \colhead{mean SSFR} }
\startdata
$\Delta z \leq 0.05$    &  minor-faint &    11.7 $\pm$ 2.1 \\
                    &  minor-bright&    2.0  $\pm$ 0.1  \\
                    &  major        &   4.4  $\pm$ 0.1  \\\cline{1-3}
$\Delta z > 0.05$       &  minor-faint  &   11.4 $\pm$ 0.8 \\
                    &  minor-bright&    1.9  $\pm$ 0.1 \\
                    &  major        &   3.8  $\pm$ 0.1 \\
\enddata
\end{deluxetable}

\begin{deluxetable}{lccc}   % Table 5
\tablecaption{Result of Spearman rank correlation test on $L_{IR}/L_{z}$ and $\Delta r_{p}$}\label{tbl5}
\tablewidth{0pt}
\tablehead{\colhead{CP} & \colhead{$r_{s}$} & \colhead{$P_{s}$} & \colhead{N} }
\startdata
$\Delta z \leq 0.05$                           &  -0.33          &  0.98           &   49 \\
$\Delta z > 0.05 $                             &  -0.17          &  0.95           &   132\\\hline
$0.05 < \Delta z\leq 0.10$                      &  -0.14          &  0.50           &   24 \\
$0.10 < \Delta z\leq 0.20 $                     &  -0.04          &  0.22           &   57 \\
$\Delta z > 0.20$                               &  -0.24          &  0.98           &   51 \\
\enddata
\tablecomments{ The parameters $r_{s}$ and $P_{s}$ are the correlation coefficient and the probability of the correlation for Spearman rank correlation test. The parameter N represents the number of CPs inside each bin of $\Delta z$.}
\end{deluxetable}

\begin{deluxetable}{llcccc}     % Table 6
\tablecaption{Comparison between the mergers in the cluster and in the field}\label{tbl9}
\tablewidth{0pt}
\tablehead{ &  \colhead{Environment}  & \colhead{Wet merger}   & \colhead{Dry merger}   & \colhead{Mix merger}  & \colhead{Dry-to-wet ratio} }
\startdata
       &  cluster           &   4        &    1     &  5        & 0.25  \\
class 1 &  field             &    143        &  4           &  21           & 0.03  \\
       &fraction in cluster & 0.03$\pm$0.01 & 0.29$\pm$0.16& 0.21$\pm$0.08&         \\ \hline
       &  cluster           &   4        &    3     &   3           & 0.75  \\
class 2 &  field             &    139        &  12      &  17           & 0.09  \\
       & fraction in cluster    & 0.03$\pm$0.02 & 0.24$\pm$0.10 & 0.18$\pm$0.08 &   \\
\enddata
\tablecomments{ In class 1, the galaxies with the $L_{IR}/L_{z}$ upper limit lower and higher than 0.5 are classified into red galaxies and blue galaxies respectively. In class 2, all galaxies without 24~$\mu$m detection are classified into red galaxies.}
\end{deluxetable}

\begin{deluxetable}{lccc}       % Table 7
\tablecaption{Comparison for wet major
mergers in clusters and in field}\label{tbl10}
\tablewidth{0pt}
\tablehead{ & \colhead{number}  & \colhead{mean SSFR}   &
\colhead{mean separation (kpc)} } \startdata
cluster         &   4          &  2.28$\pm$0.42 &   13.29$\pm$0.74  \\
field          &    109        &  4.36$\pm$0.04 &   13.14$\pm$0.07  \\
\enddata
\end{deluxetable}

\clearpage
%%Figures

\begin{figure}                    % Figure 1
\plotone{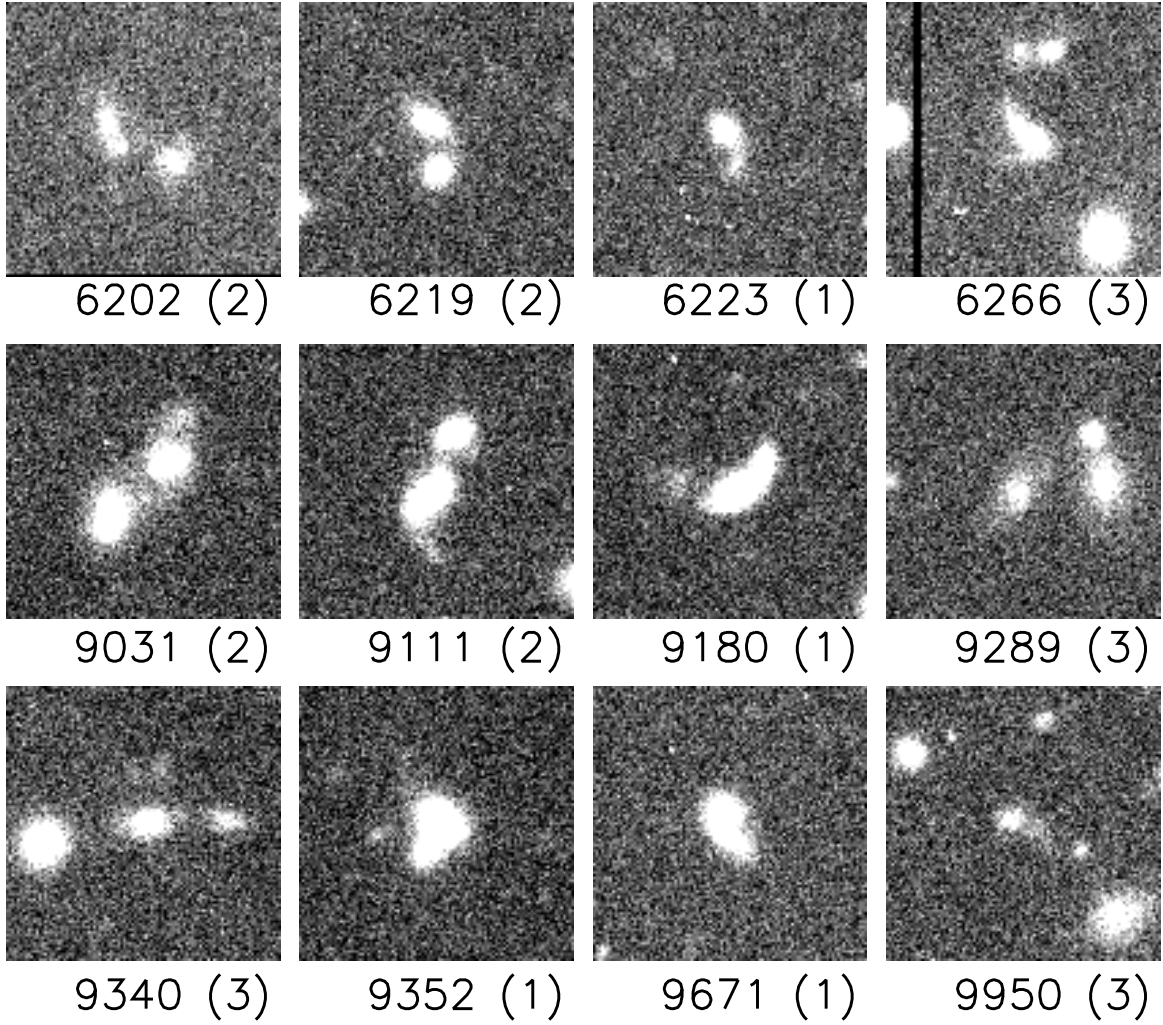}
\caption{Examples of merging systems without contamination of stars. The number below each image
is its original ID in the merging galaxy catalog;
the number in parenthesis represents its classification in this study: 1 for MG, 2 for CP, 3 for CM. }
  \label{fig1}
\end{figure}

\clearpage

\begin{figure}                   % Figure 2
\plotone{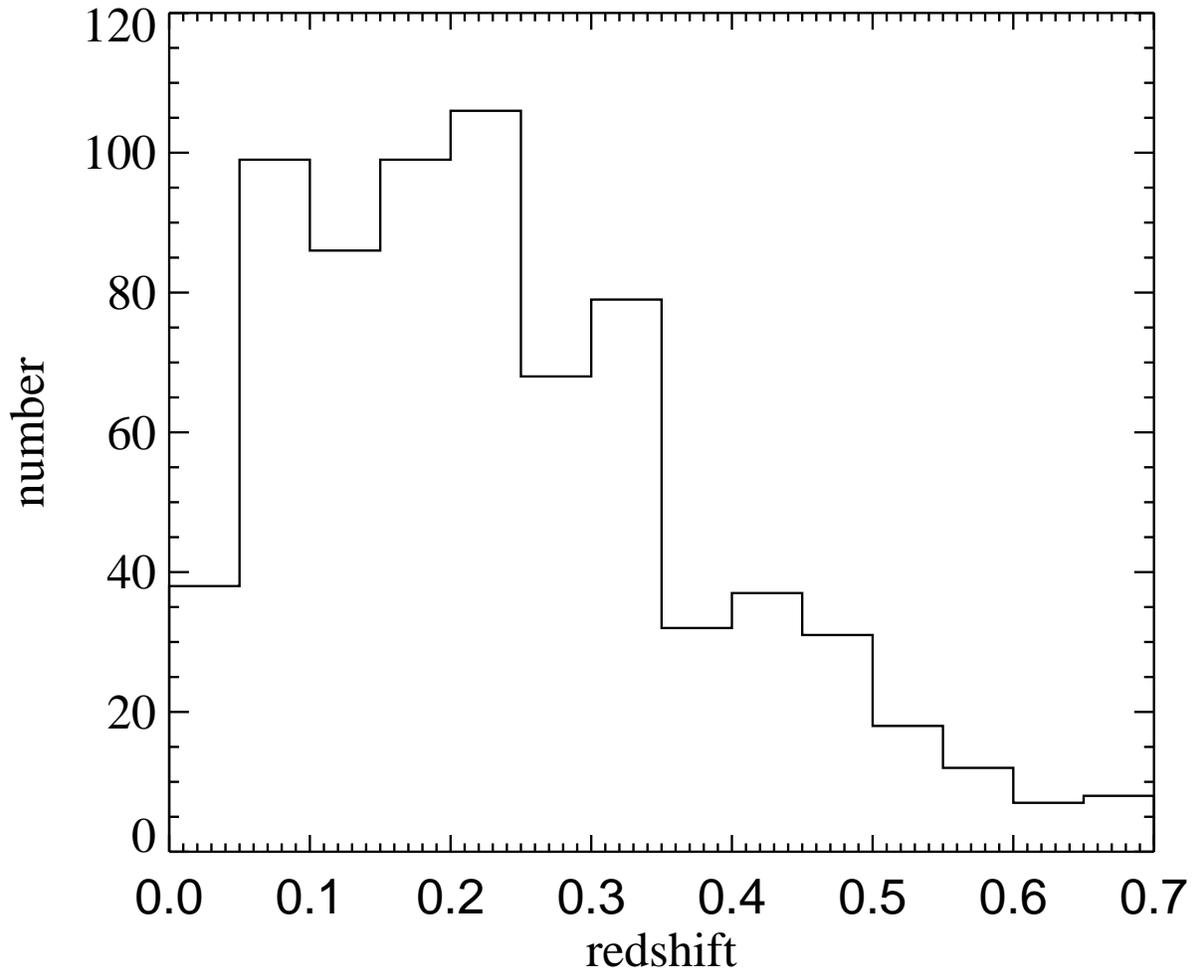}
\caption{Redshift distribution of merging galaxies.}
\label{fig2}
\end{figure}

\clearpage

\begin{figure}                   % Figure 3
\plotone{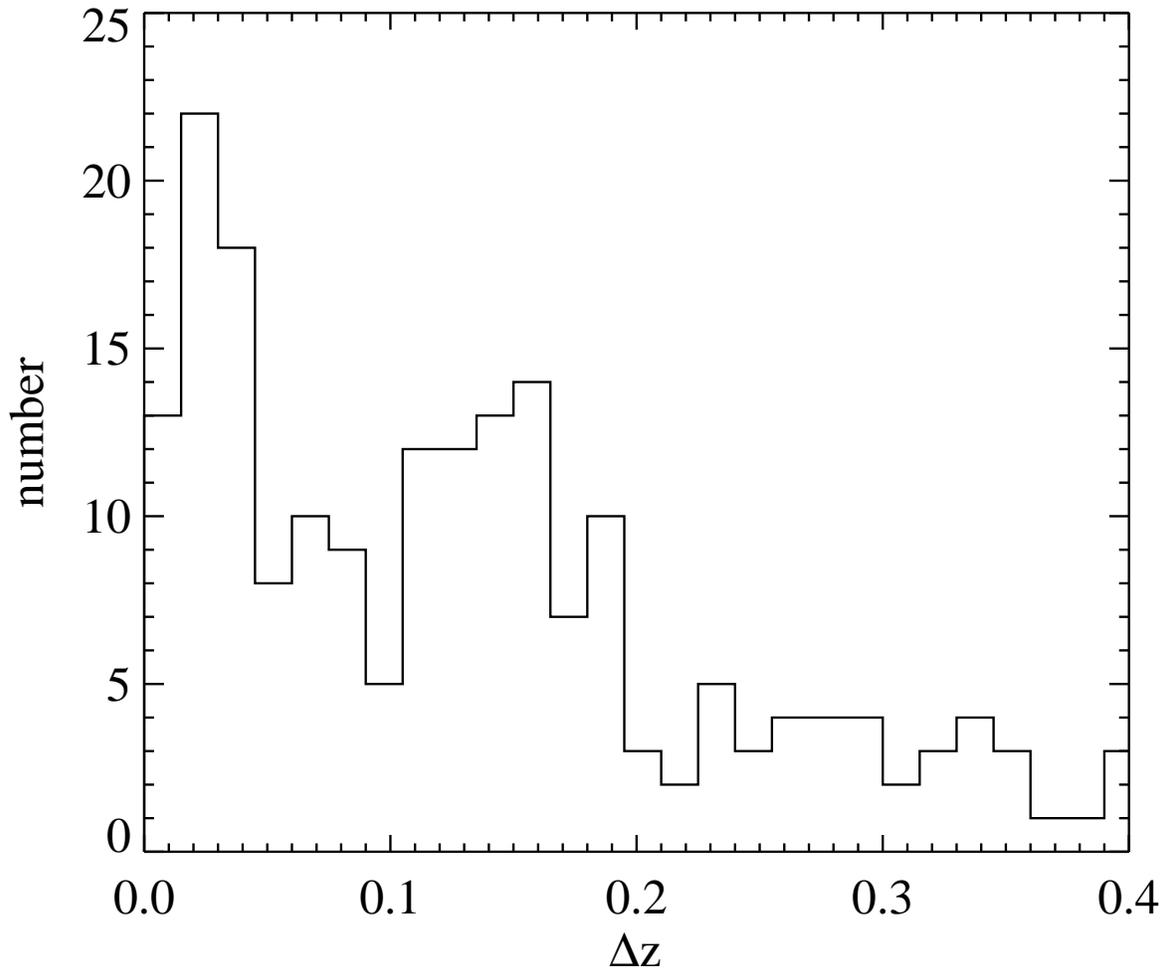}
\caption{Distribution of redshift difference for galaxy pairs.}
  \label{fig3}
\end{figure}

\clearpage

\begin{figure}                   % Figure 4
\plotone{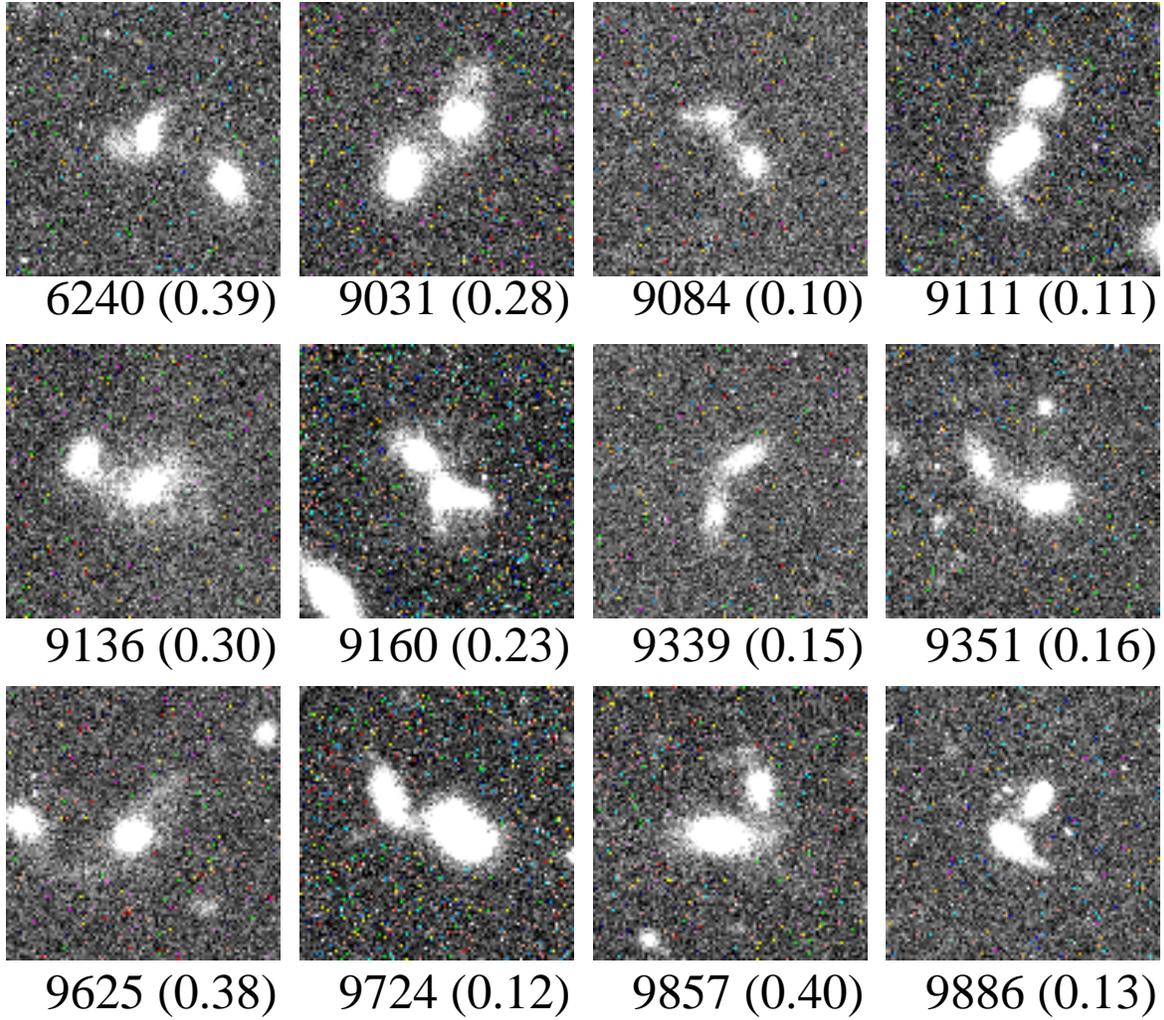}
\caption{Examples of CPs with $\Delta z >$ 0.05. The number below each image is its original ID in the merging galaxy catalog; the number in parenthesis represents the redshift difference between its member galaxies. }
\label{fig4}
\end{figure}

\clearpage

\begin{figure}                   % Figure 5
\plotone{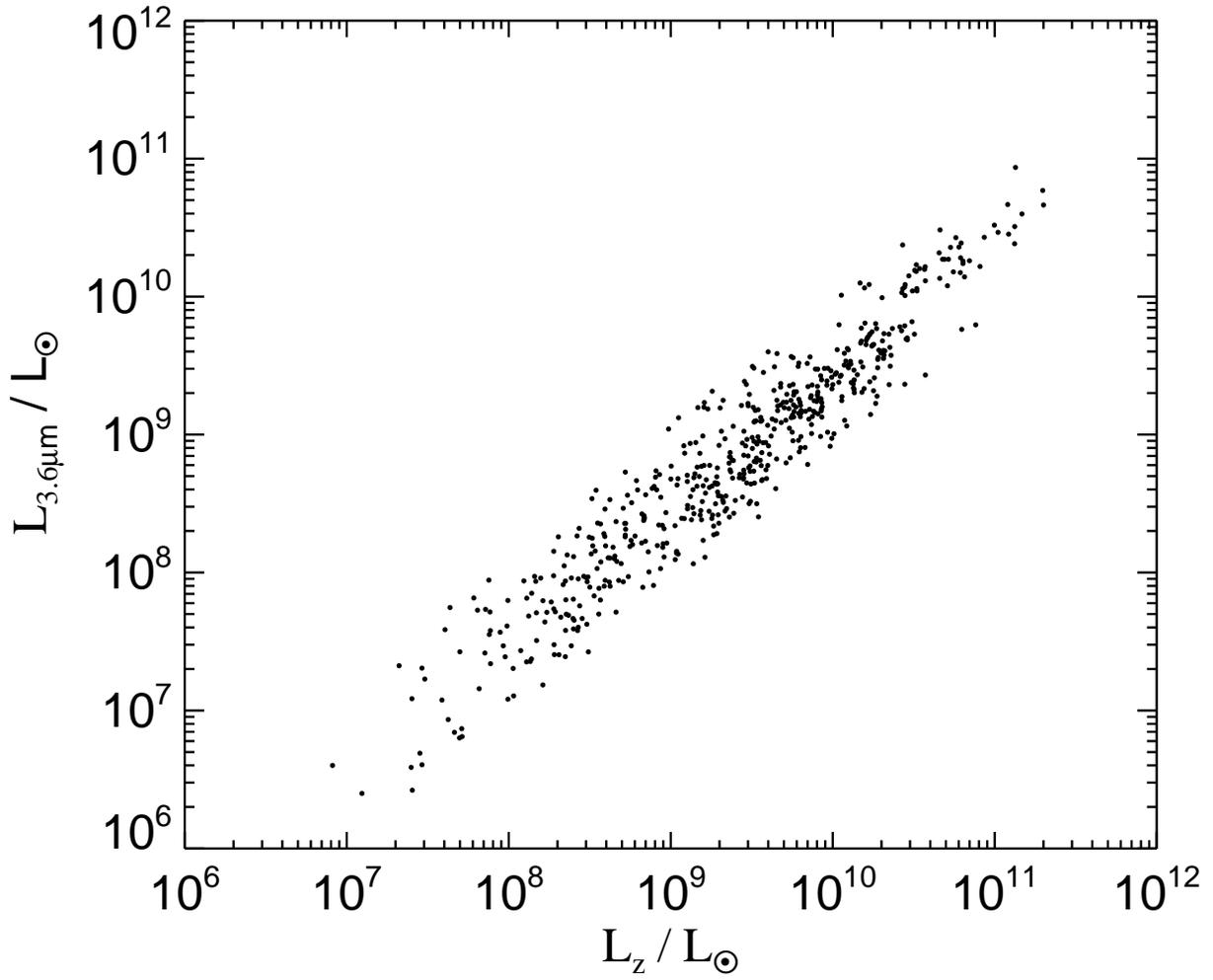}
\caption{Plots of the 3.6~$\mu$m versus the $z$-band luminosity of galaxies.}
\label{fig5}
\end{figure}

\clearpage

\begin{figure}                   % Figure 6
\begin{center}
\includegraphics[]{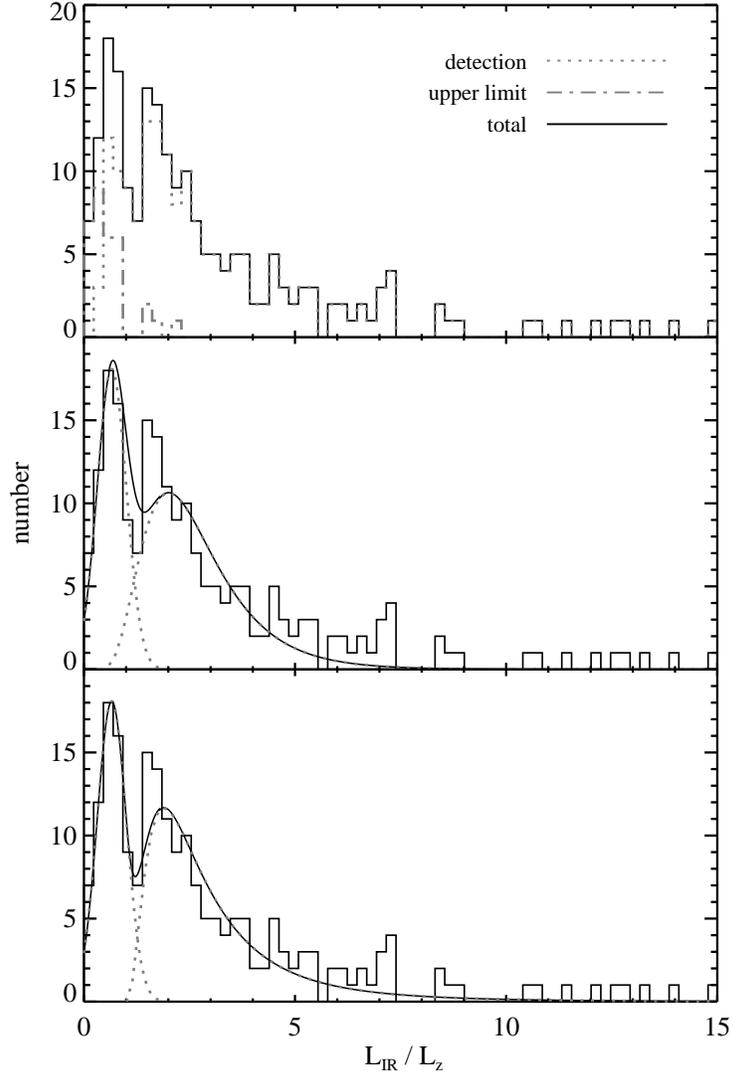}
\caption{SSFR distribution of galaxy pairs. Middle panel shows a combined model of normal
and lognormal fits; and the bottom panel shows a combination of normal and
modified-lognormal fits. In the top panel, dotted and
dash dot lines represent the CPs with and without detections at 24~$\mu$m
respectively, and the solid line represents the overall
sample.}
\label{fig6}
\end{center}
\end{figure}

\clearpage

\begin{figure}                   % Figure 7
\plotone{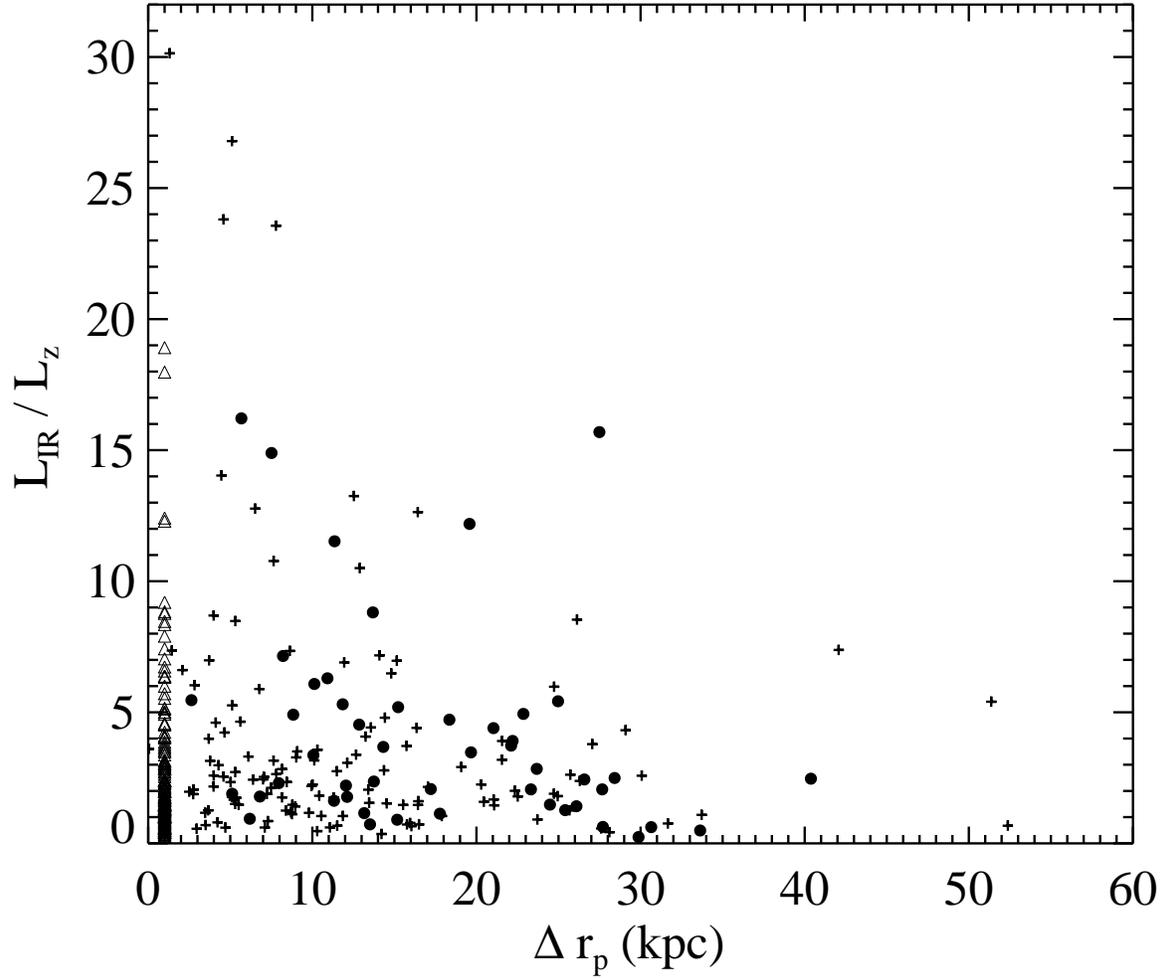}
\caption{$L_{IR}/L_{z}$ as a function of projected separation for CPs with $\Delta z \leq 0.05$ (solid circle), $\Delta z \geq 0.05$ (plus), and MGs (triangle).}
\label{fig7}
\end{figure}

\clearpage

\begin{figure}                   % Figure 8
\plotone{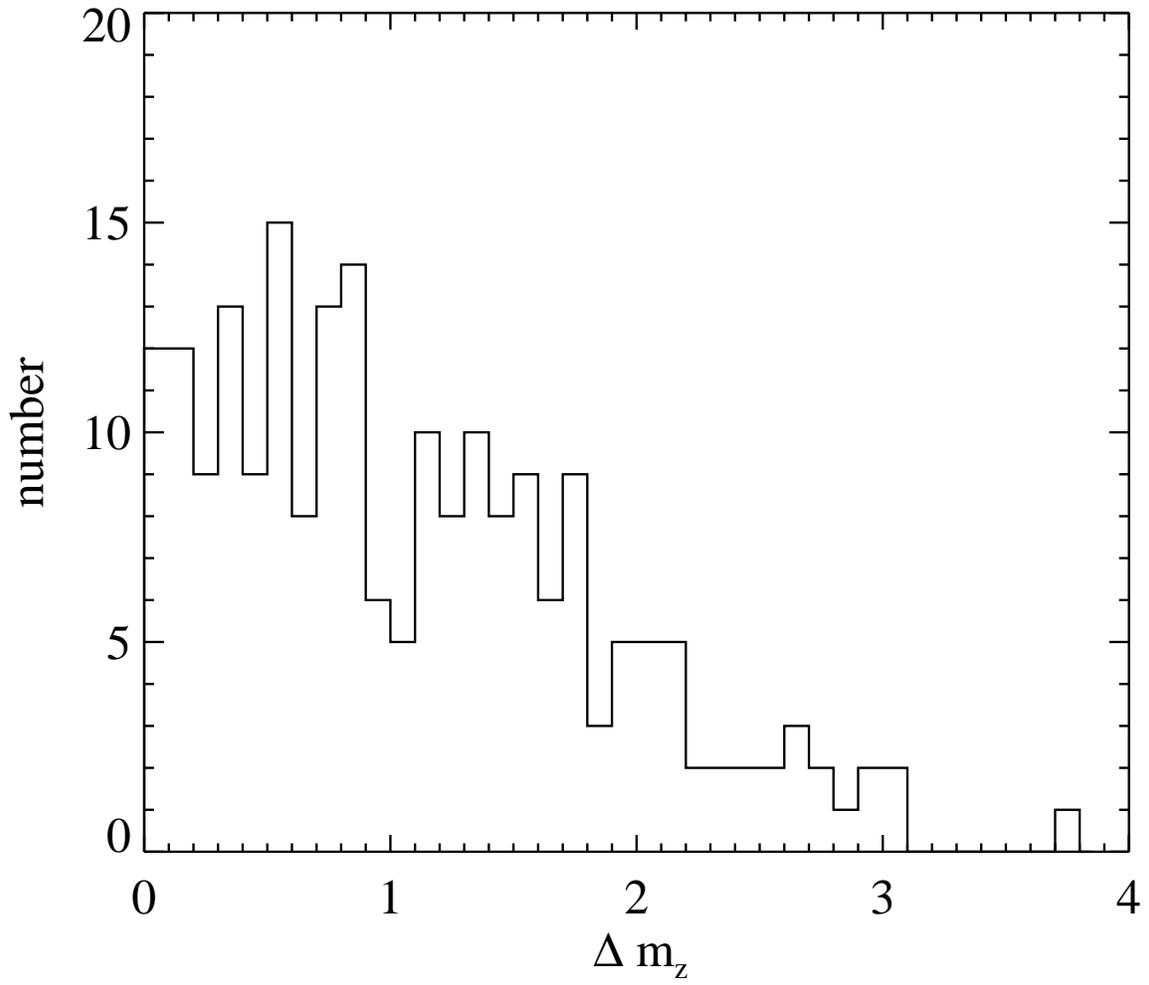}
\caption{Distribution of magnitude difference for CPs.}
  \label{fig8}
\end{figure}

\clearpage

\begin{figure}                   % Figure 9
 \plottwo{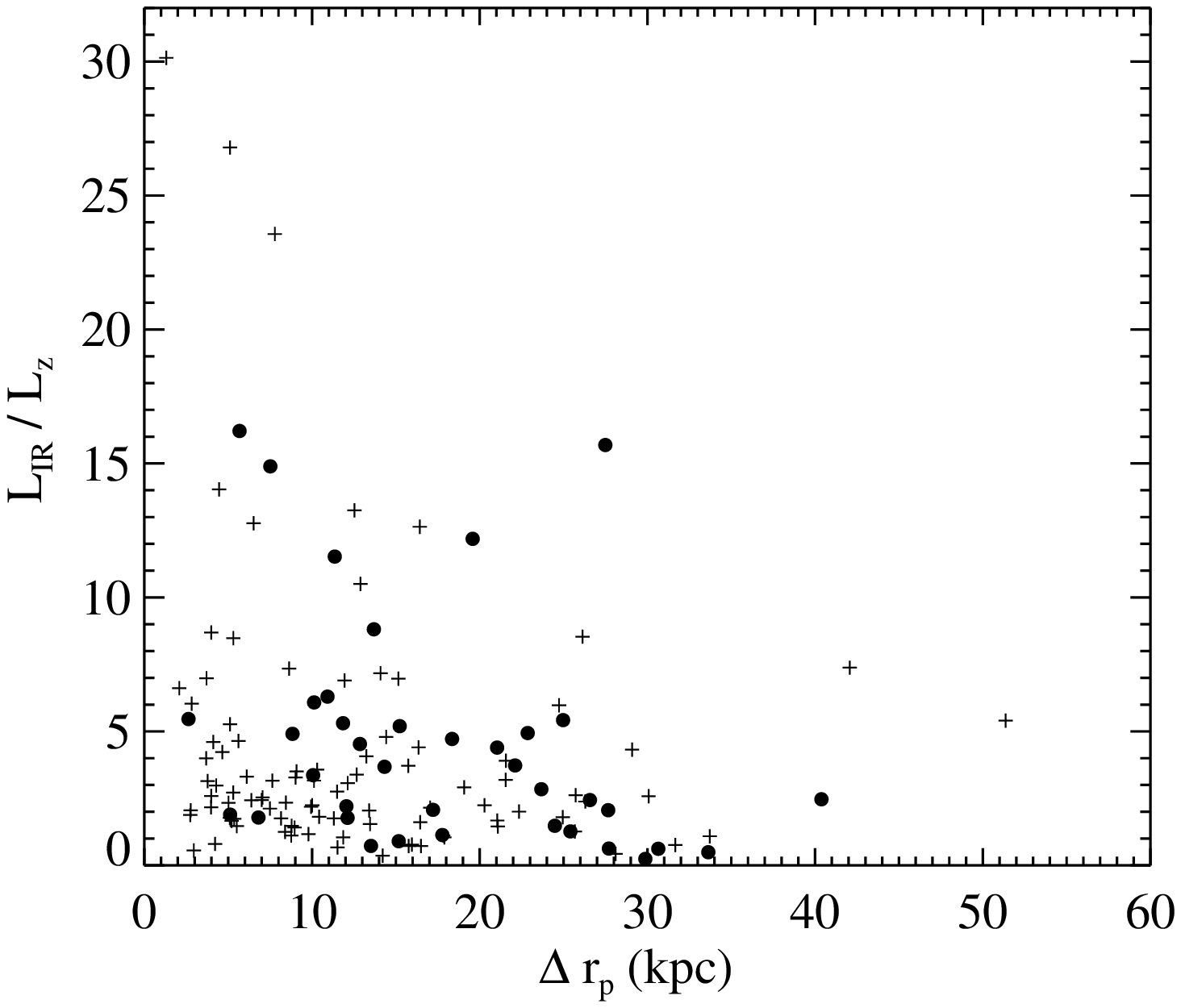}{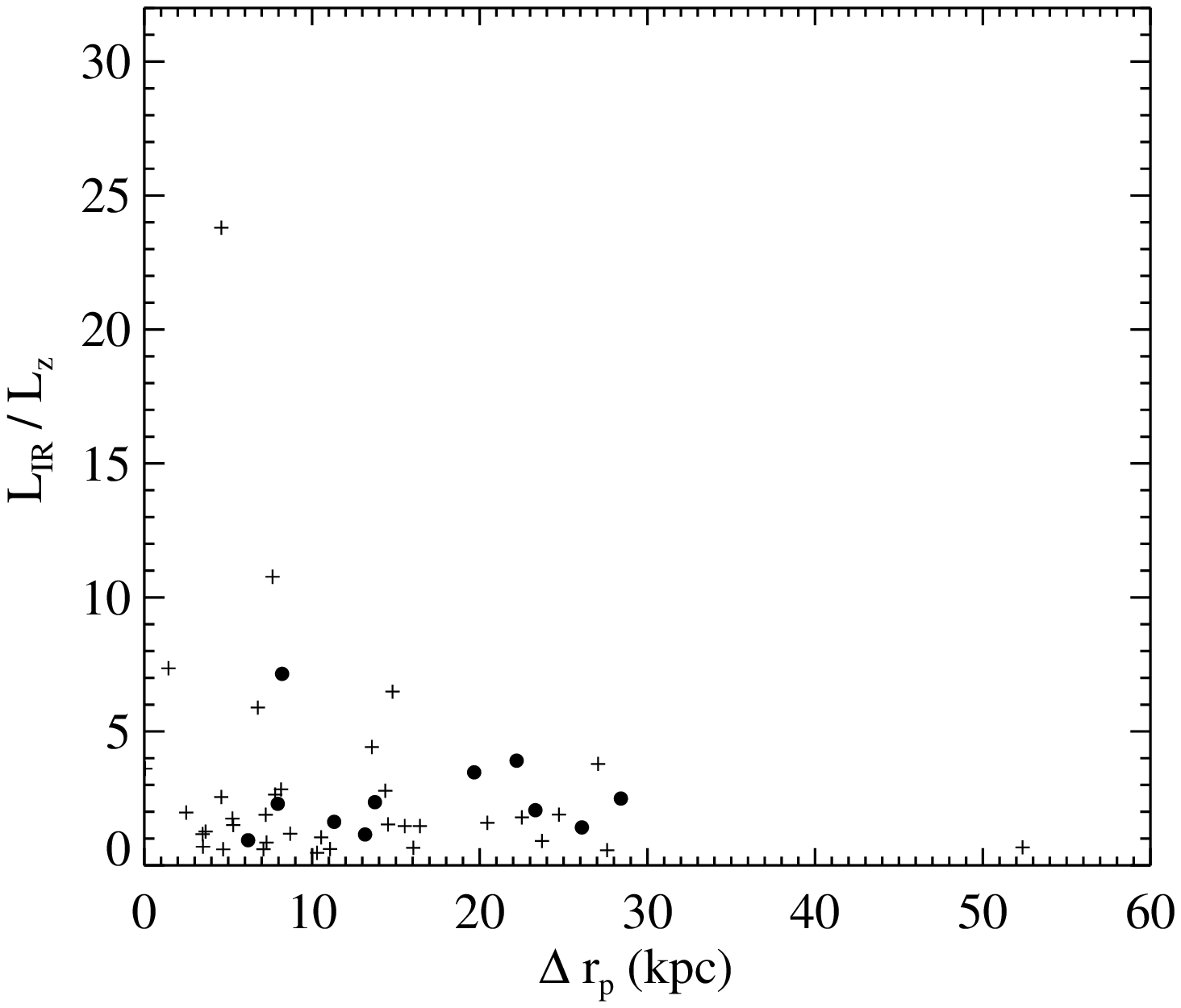}
 \caption{$L_{IR}/L_{z}$ versus $\Delta r_{p}$ for CPs with $\Delta m_{z} < 1.5$ (left) and for CPs with  $\Delta m_{z} \geq 1.5$ (right). Solid circles are the CPs with $\Delta z \leq$ 0.05; pluses are the CPs with $\Delta z >$ 0.05.}
 \label{fig9}
\end{figure}

\clearpage

\begin{figure}                   % Figure 10
 \plotone{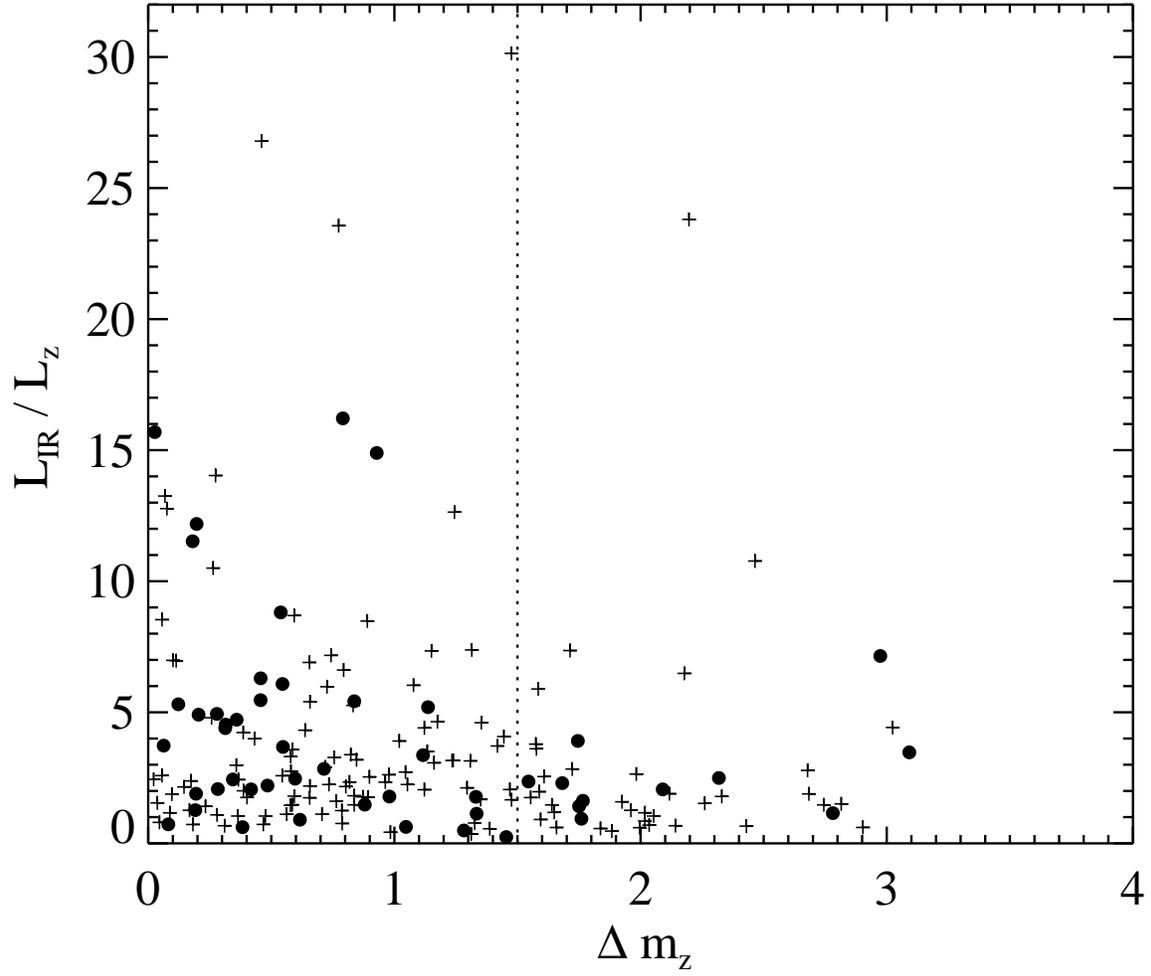}
\caption{$L_{IR}/L_{z}$ as a function of $\Delta m_{z}$ for the CPs with $\Delta z \leq$ 0.05 (solid circle) and with $\Delta z >$ 0.05 (plus). The dotted line denotes a dividing criterion between the major and the minor mergers.}
  \label{fig10}
\end{figure}

\clearpage

\begin{figure}                   % Figure 11
 \plottwo{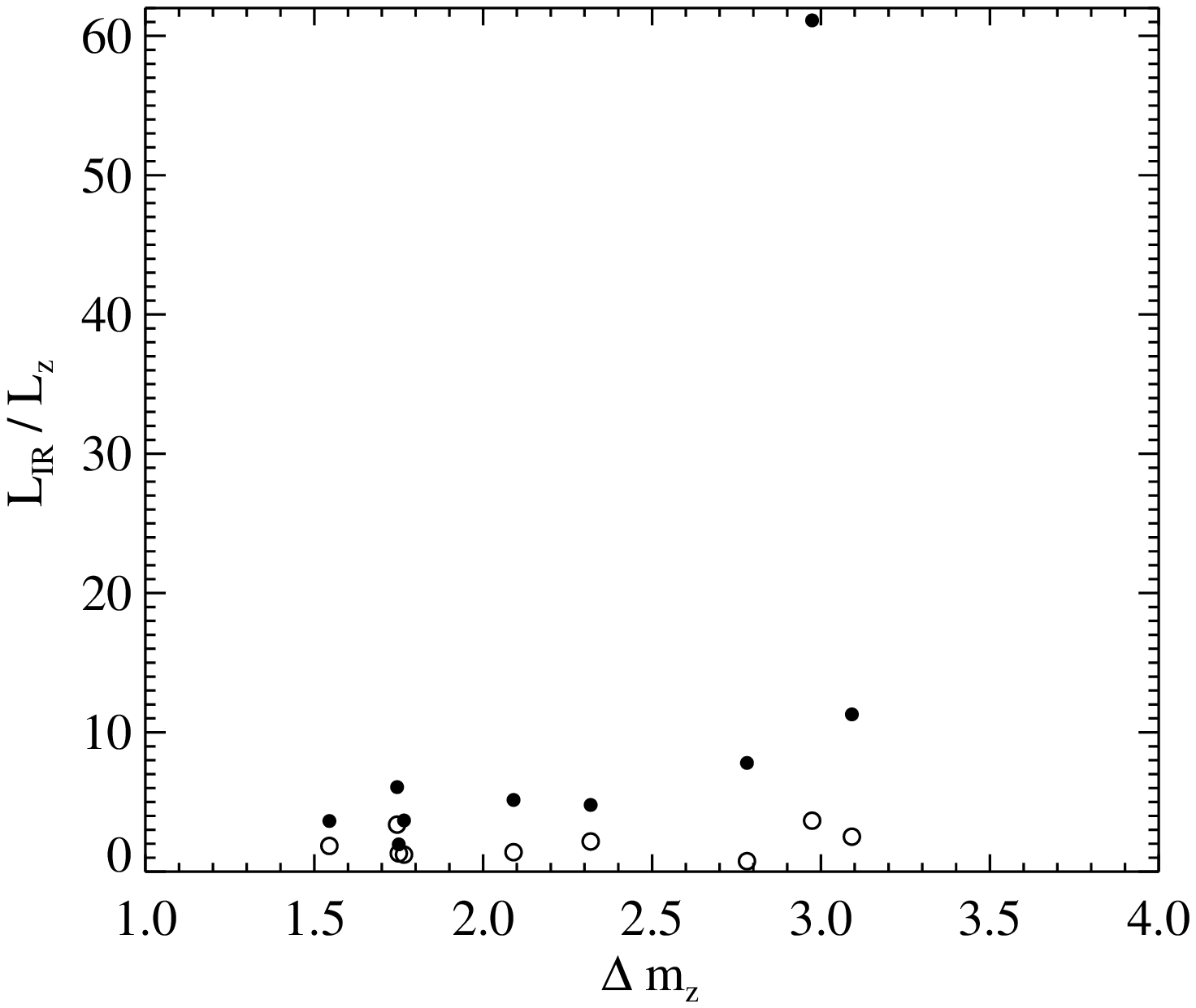}{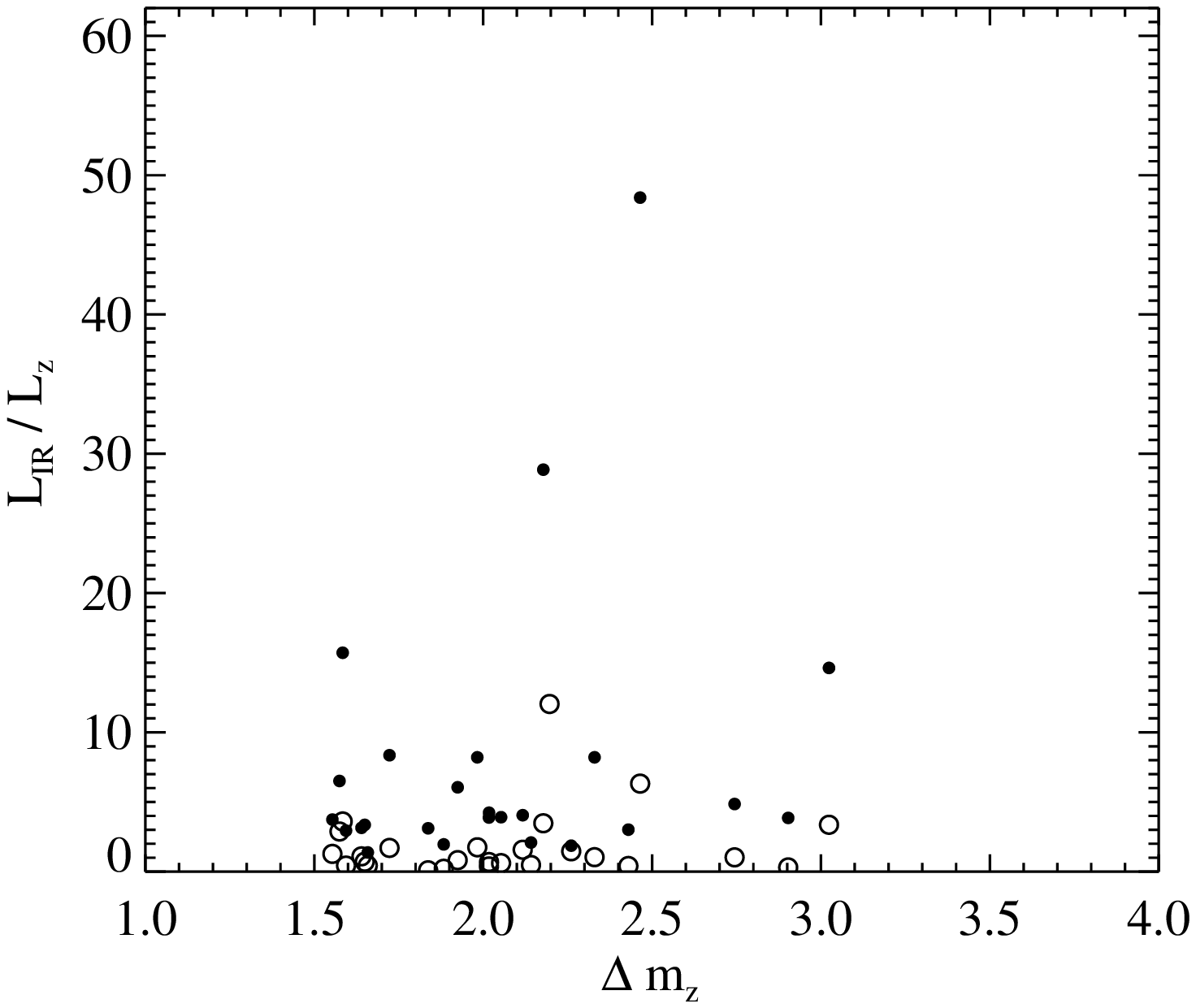}
 \caption{$L_{IR}/L_{z}$ as a function of $\Delta m_{z}$ for the minor mergers with $\Delta z \leq 0.05$ (left) and those with $\Delta z > 0.05$ (right). Solid circles are the faint member galaxies; open circles are the bright member galaxies. }
 \label{fig11}
\end{figure}

\begin{figure}                   % Figure 12
 \plotone{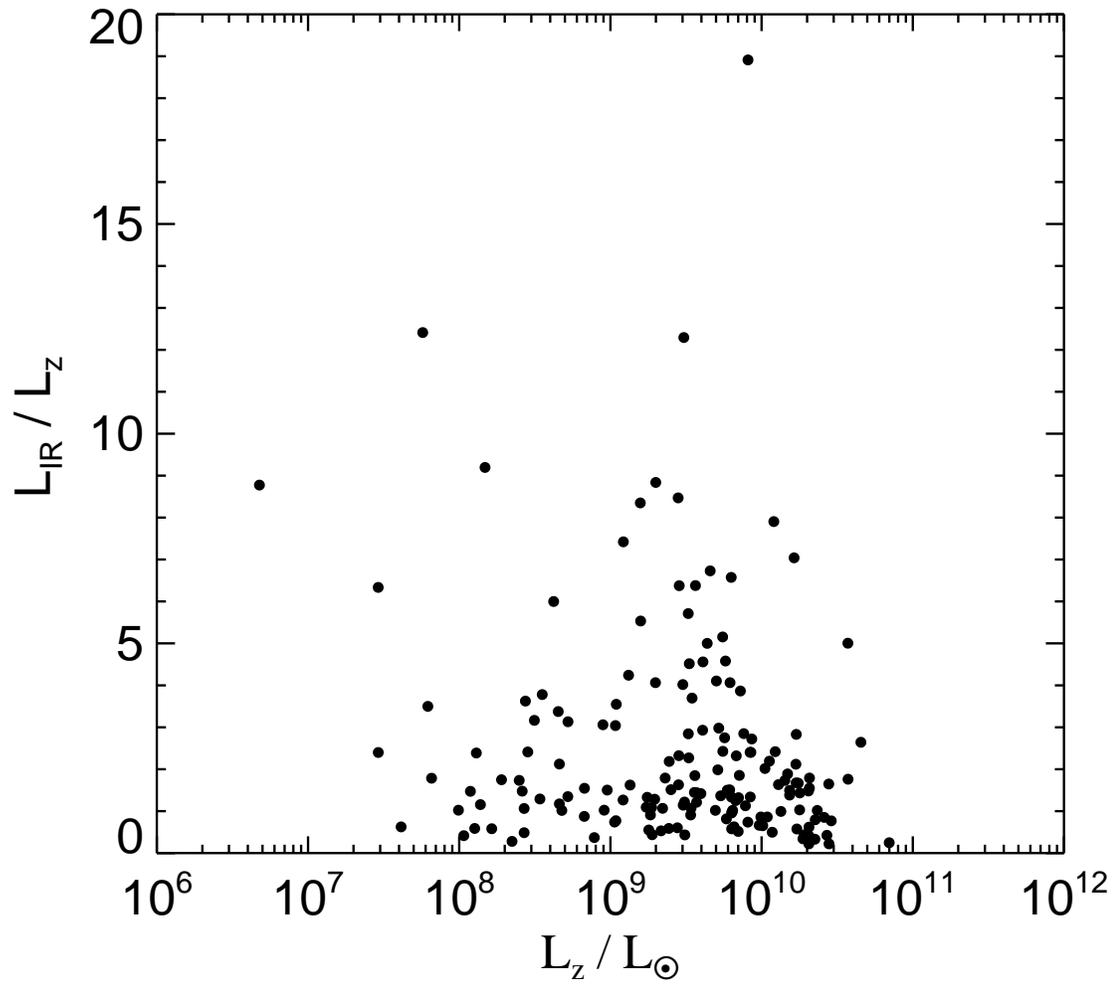}
\caption{$L_{IR}/L_{z}$ as a function of $L_{z}$ for the MGs in the field.}
  \label{fig12}
\end{figure}


\begin{thebibliography}{}

\bibitem[Abadi et al.(1999)]{aba99}
    Abadi, M. G., Moore, B., \& Bower, R. G. 1999, \mnras, 308, 947
\bibitem[Alonso et al.(2004)]{alo04}
    Alonso M. S., et al. 2004, \mnras, 352, 1081
\bibitem[Barnes \& Hernquist et al.(1996)]{bar96}
    Barnes, J., \& Hernquist, L., et al. 1996, \apj, 471, 115
\bibitem[Barton et al.(2000)]{bar00}
    Barton, E. J., et al. 2000, \apj, 530, 660
\bibitem[Bekki et al.(2002)]{bek02}
    Bekki, K., Couch, W. J., \& Shioya, Y. 2002, \apj, 577, 651
\bibitem[Blanton et al.(2005)]{bla05}
    Blanton, M. R., et al. 2005, \apj, 631, 208
\bibitem[Charlot(1996)]{cha96}
    Charlot, S. 1996, in The Universe at Hight-z, Large-Scale Structure and the Cosmic Microwave Background, ed. E. Martinez-Gonzalez \& J. L. Sanz (Berlin: Springer), 53
\bibitem[Chary \& Elbaz(2001)]{cha01}
    Chary, R., \& Elbaz, D. 2001, \apj, 556, 562
\bibitem[Davoodi et al.(2006)]{dav06}
    Davoodi, P., et al. 2006, \mnras, 371, 1113
\bibitem[Dale et al.(2001)]{dal01}
    Dale, D. A., et al. 2001, \apj, 549, 215
\bibitem[Dale \& Helou(2002)]{dal02}
    Dale, D. A., \& Helou, G. 2002, \apj, 576, 159
\bibitem[de Ravel et al.(2009)]{rav09}
    de Ravel, L., et al. 2009, \aap, 498, 379
\bibitem[Dressler(1980)]{dre80}
    Dressler, A. 1980, \apj, 236, 351
\bibitem[Elbaz et al.(2002)]{elb02}
    Elbaz, D., et al. 2002, \aap, 384, 848
\bibitem[Fujita et al.(1999)]{fuj99}
    Fujita, Y., \& Nagashima, M. 1999, \apj, 516, 619
\bibitem[Gunn \& Gott(1972)]{gun72}
    Gunn, J. E., \& Gott, J. R. I. 1972, \apj, 176, 1
\bibitem[Hancock et al.(2007)]{han07}
     Hancock, M., et al. 2007, \aj, 133, 676
\bibitem[Hibbard \& van Gorkom(1996)]{hib96}
     Hibbard, J. E., \& van Gorkom, J. H. 1996, \aj, 111, 655
\bibitem[Hwang \& Chang(2009)]{hwa09}
    Hwang, C. Y., \& Chang, M. Y. 2009, \apjs, 181, 233
\bibitem[Kennicutt(1998)]{ken98}
    Kennicutt, R. C., Jr. 1998, \apj, 498, 541
\bibitem[Koester et al.(2007)]{koe07}
    Koester, B. P., et al. 2007, \apj, 660, 239
\bibitem[Kron(1980)]{kro80}
    Kron, R. G. 1980, \apjs, 43, 305
\bibitem[Lagache et al.(2003)]{lag03}
    Lagache, G., Dole, H., \& Puget, J.-L. 2003, \mnras, 338, 555
\bibitem[Lambas et al.(2003)]{lam03}
    Lambas, D. G., et al. 2003, \mnras, 346, 1189
\bibitem[Larson et al.(1980)]{lar80}
    Larson, R. B., Tinsley, B. M., \& Caldwell, C. N. 1980, \apj, 237, 692
\bibitem[Li et al.(2007)]{li07}
    Li, H.-N., Wu, H., Cao, C., \& Zhu, Y.-N. 2007, \aj, 134, 1315
\bibitem[Lin et al.(2007)]{lin07}
    Lin, L., et al. 2007, \apj, 660, L51
\bibitem[Lin et al.(2008)]{lin08}
    Lin, L., et al. 2008, \apj, 681, 232
\bibitem[Lonsdale et al.(2003)]{lon03}
    Lonsdale, C. J., et al. 2003, \pasp, 115, 897
\bibitem[Lonsdale et al.(2004)]{lon04}
    Lonsdale, C. J., et al. 2004, \apjs, 154, 54
\bibitem[Madau et al.(1998)]{mad98}
    Madau, P., Pozzetti, L., \& Dickinson, M. 1998, \apj, 498, 106
\bibitem[Mihos \& Hernquist(1996)]{mih96}
    Mihos, J. C., \& Hernquist, L. 1996, \apj, 464, 641
\bibitem[Nikolic et al.(2004)]{nik04}
    Nikolic, B., Cullen, H., \& Alexander, P. 2004, \mnras, 355, 874
\bibitem[Nulsen et al.(1982)]{nul82}
    Nulsen, P. E. J. 1982, \mnras, 198, 1007
\bibitem[Papovich \& Bell(2002)]{pap02}
    Papovich, C., \& Bell, E. F. 2002, \apj, 579, L1
\bibitem[Polletta et al.(2007)]{pol07}
    Polletta, M., et al. 2007, \apj, 663, 81
\bibitem[Rowan-Robinson et al.(2008)]{RR08}
    Rowan-Robinson, M., et al. 2008, \mnras, 386, 697
\bibitem[Schweizer(1982)]{sch82}
    Schweizer, F. 1982, \apj, 252, 455
\bibitem[Takeuchi et al.(2005)]{tak05}
    Takeuchi, T. T., et al. 2005, \aap, 432, 423
\bibitem[Tissera et al.(2002)]{tis02}
    Tissera, P. B., et al. 2002, \mnras, 333, 327
\bibitem[Toomre \& Toomre(1972)]{too72}
    Toomre, A. \& Toomre, J. 1972, \apj, 178, 623
\bibitem[Woods et al.(2006)]{woo06}
    Woods, D. F., et al. 2006, \aj, 132, 197
\bibitem[Woods \& Geller(2007)]{woo07}
    Woods, D. F. \& Geller, M. J. 2007, \aj, 134, 527
\bibitem[Wu et al.(2005)]{wu05}
    Wu, H., et al. 2005, \apj, 632, L79
\bibitem[York et al.(2000)]{yor00}
    York, D. G., et al. 2000, \aj, 120, 1579
\end{thebibliography}
\end{document}